\documentclass[aps,prd,amsmath,amssymb,preprint,superscriptaddress,longbibliography]{revtex4-1}

\usepackage[utf8]{inputenc}
\usepackage[T1]{fontenc}
\usepackage{graphicx}
\usepackage[dvipsnames]{xcolor}
\usepackage{dcolumn}
\usepackage{bm}
\usepackage[colorlinks=true, linkcolor=RoyalBlue, citecolor=RoyalBlue]{hyperref}
\usepackage[per-mode=symbol, separate-uncertainty]{siunitx}
\usepackage[capitalise]{cleveref}
\usepackage{enumerate}
\usepackage{float}
\usepackage{lineno}

\usepackage[normalem]{ulem} 

\definecolor{ablue}{rgb}{0,0,0}

\begin{document}
\title{Exciton transport driven by spin excitations in an antiferromagnet} 	

\author{Florian Dirnberger$^\ddagger$}
\email{f.dirnberger@tum.de}
\affiliation{Institute of Applied Physics and Würzburg-Dresden Cluster of Excellence ct.qmat, TUD Dresden University of Technology, Dresden, 01187, Germany}
\affiliation{Zentrum für QuantumEngineering (ZQE), Technical University of Munich, Garching, Germany}
\affiliation{Department of Physics, TUM School of Natural Sciences, Technical University of Munich, Munich, Germany}
\thanks{Authors contributed equally.}	
	
\author{Sophia Terres$^\ddagger$}
\affiliation{Institute of Applied Physics and Würzburg-Dresden Cluster of Excellence ct.qmat, TUD Dresden University of Technology, Dresden, 01187, Germany}
\thanks{Authors contributed equally.}	

\author{Zakhar A. Iakovlev}
\affiliation{Ioffe Institute, 194021 St. Petersburg, Russia}
	
\author{Kseniia Mosina}
\author{Zdenek Sofer}
\affiliation{Department of Inorganic Chemistry, University of Chemistry and Technology Prague, Technickaá 5, 166 28 Prague 6, Czech Republic}
	
\author{Akashdeep Kamra}
\affiliation{Department of Physics and Research Center OPTIMAS, Rheinland-Pf\"alzische Technische Universit\"at Kaiserslautern-Landau, 67663 Kaiserslautern, Germany}
\affiliation{Departamento de Física Teórica de la Materia Condensada and Condensed Matter Physics Center (IFIMAC), Universidad Autónoma de Madrid, E- 28049 Madrid, Spain}
	
\author{Mikhail M. Glazov}
\affiliation{Ioffe Institute, 194021 St. Petersburg, Russia}
	
\author{Alexey Chernikov}
\email{alexey.chernikov@tu-dresden.de}
\affiliation{Institute of Applied Physics and Würzburg-Dresden Cluster of Excellence ct.qmat, TUD Dresden University of Technology, Dresden, 01187, Germany}
	
\begin{abstract}
A new class of optical quasiparticles called magnetic excitons recently emerged in magnetic van der Waals materials~\cite{Seyler2018,Zhang2019,Kang2020,Wilson2021}.
Akin to the highly effective strategies developed for electrons~\cite{Baibich1988,Binasch1989}, the strong interactions of these excitons with the spin degree of freedom~\cite{Seyler2018,Wilson2021,Bae2022,Diederich2022,Dirnberger2023} may provide innovative solutions for long-standing challenges in optics, such as steering the flow of energy and information~\cite{Unuchek2018,Dong2021,Jakhangirkhodja2023}. 
Here, we demonstrate transport of excitons by spin excitations in the van der Waals antiferromagnetic semiconductor CrSBr.  
Key results of our study are the observations of ultrafast, nearly isotropic exciton propagation substantially enhanced at the Néel temperature, transient contraction and expansion of the exciton clouds at low temperatures, as well as superdiffusive behavior in bilayer samples.
These signatures largely defy description by commonly known exciton transport mechanisms and are related to the currents of incoherent magnons induced by laser excitation instead.
We propose that the drag forces exerted by these currents can effectively imprint characteristic properties of spin excitations onto the motion of excitons.
The universal nature of the underlying exciton-magnon scattering promises driving of excitons by magnons in other magnetic semiconductors and even in non-magnetic materials by proximity in heterostructures, merging the rich physics of magneto-transport with optics and photonics.
\end{abstract}

\maketitle

More than three decades ago, the giant magnetoresistance effect~\cite{Baibich1988,Binasch1989} demonstrated the extensive potential of controlling electrons using the spin degree of freedom in solids.
The profound impact of this discovery on science and technology spawned the field of spintronics and ultimately came to play an important role in modern electronics.
Now, reports of excitons in magnetic van der Waals crystals~\cite{Seyler2018,Zhang2019,Kang2020,Wilson2021,Birowska2021,Kim2022} and their interactions with magnetic spin order raise the question whether similar developments are on the brink of transforming optics and photonics. 
High-speed propagation, anomalous dispersion, exceptional coherence and thermopower~\cite{Barman2021,Lebrun2018,Tu2020,Lee2021,Hortensius2021,Wei2022} can be extremely attractive features of spin excitations (magnons and para-magnons) in this context.
This promise, however, rests on the expectation that the recently reported coupling of excitons and magnons~\cite{Bae2022,Diederich2022,Dirnberger2023,Sun2024} can indeed be leveraged to control optical quasiparticles in solids. 

\begin{figure*}[]
	\includegraphics[scale=0.7]{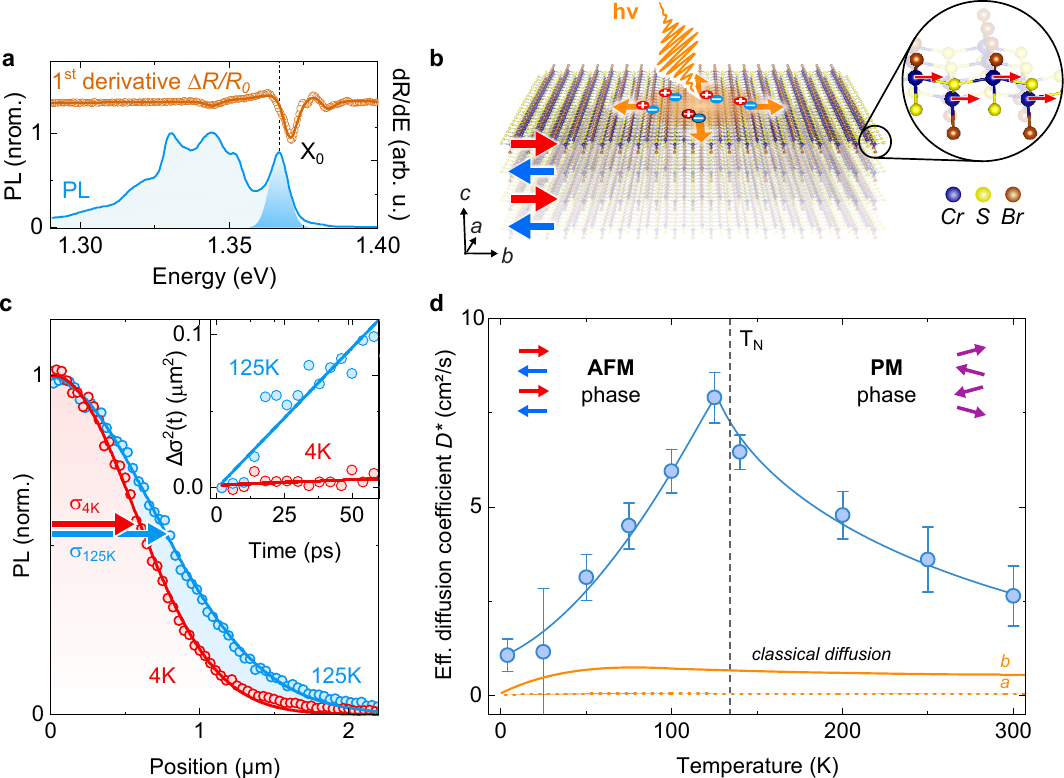}
	\caption{\textbf{Exciton transport in CrSBr across the AFM-PM magnetic phase transition.} 
		\textbf{a}~PL spectrum and derivative of reflectance contrast (with a fit curve, see Methods) recorded at $T=4$~K. 
		Dashed line and the Gaussian profile mark the optical signature of the fundamental exciton $X_0$ transition in CrSBr~\cite{Wilson2021,Meineke2024}. 
		\textbf{b}~Schematic illustration of the CrSBr structure in the AFM phase and the optical injection of excitons in the experiments. 
		\textbf{c}~PL cross-section profiles along $b$ measured at 4 and 125\,K. 
		The PL is integrated in energy and in time from 0 to 150\,ps after pulsed excitation. 
		Arrows represent the standard deviation $\sigma$ extracted from Gaussian fits. 
		Inset: Corresponding time dependence of the relative mean squared displacement $\Delta\sigma^2(t)$. 
		Solid lines are linear fits to the data. 
		\textbf{d}~Blue circles: Temperature dependence of the effective diffusion coefficient $D^*$, extracted by evaluating $D^{*}=\frac{1}{2}\times\partial \Delta \sigma^2/\partial t$ over the first 80\,ps along the $b$--axis.
		Blue line is a guide to the eye. 
		Solid and dotted orange lines show the expected classical diffusion of excitons along the $a$-- and $b$--axis. 
		Labels and the black dashed line mark the Néel temperature and the AFM-PM phase transition, respectively. 
		Error bars indicate the statistical error of the fit. 
		Data in \textbf{c} and \textbf{d} obtained under 390\,$\mu$J/cm$^2$ fluence at 1.61\,eV excitation energy, corresponding to an exciton density in the range of few 10$^{12}$\,cm$^{-2}$ per layer.
		Similar results, obtained at smaller fluence for an excitation energy of 1.77\,eV close to the $B$ exciton resonance, are shown in Extended Data \cref{fig:Fig-R2}.}
	\label{fig:1}
\end{figure*}		
	
A key material to explore this question is the van der Waals (vdW) magnetic semiconductor CrSBr~\cite{Goser1990,Telford2020}.
At low temperatures, CrSBr exhibits strong magnetization along the in-plane $b$--axis that alternates between layers in the out-of-plane $c$--axis (cf. \cref{fig:1}b).
Moderate magnetic fields are already sufficient to switch the antiferromagnetic (AFM) ground state into a ferromagnetic (FM) configuration. 
As the temperature rises, an increasingly larger number of thermal magnons progressively suppresses long-range magnetic order until the material becomes paramagnetic (PM) above the Néel temperature at $T_N=132$\,K~\cite{Goser1990,Telford2020,Long2023}. 
A local temperature gradient generates a flux of incoherent magnons~\cite{Canetta2024}.
Most importantly, CrSBr hosts tightly bound excitons that interact strongly with light~\cite{Dirnberger2023}, are tunable by magnetic fields~\cite{Wilson2021,Klein2022-1,Tabataba2023} and couple to both coherent and incoherent magnons~\cite{Bae2022,Diederich2022,Dirnberger2023,Sun2024}.
This renders it an ideal platform to study the impact of magnon currents on the excitonic motion.

Here, we demonstrate the transport of excitons in CrSBr and present a series of experimental signatures implicating the drag of excitons by magnons.
For up to tens of picoseconds after the excitation by a short light pulse, excitons are observed to move remarkably fast.
Their propagation correlates with the magnetic phase and reaches a maximum at the Néel temperature.
Corresponding effective diffusion coefficients are as high as $\SI{150}{cm^2/s}$, exceeding expectations from classical exciton diffusion by orders of magnitude.
Moreover, for the majority of excitation conditions, the exciton propagation is quasi-isotropic in the vdW plane, in stark contrast to the highly anisotropic exciton effective masses dictated by the electronic dispersion of CrSBr.
Instead, it matches the nearly isotropic in-plane propagation of thermal magnons and their group velocities.
Finally, at low temperatures and excitation densities, we observe a complete reversal of the exciton propagation direction, from expansion to contraction, and find ultrafast, superdiffusive behavior in bilayers (2L) with effective velocities reaching 41\,km/s within the first 15\,ps.

\begin{figure*}[]
	\includegraphics[scale=0.9]{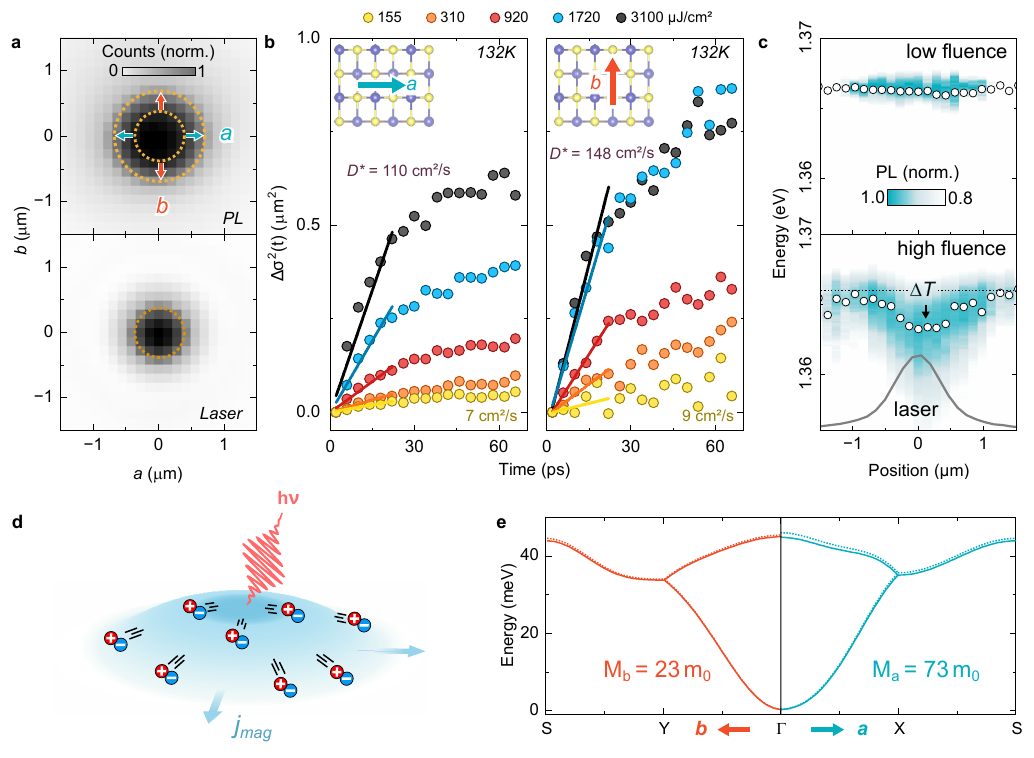}
	\caption{\textbf{Direction and fluence dependence of exciton transport in 10L.} 
		\textbf{a}~Top: Spatial PL profile recorded at $T_N=132$\,K under a fluence of \SI{3100}{\micro\J\per\square\cm}. 
		Broadening along $a$- and $b$-directions is indicated by red and blue arrows. 
		Dashed circle marks the $\sigma$ extracted from Gauss fits. Bottom: Spatial profile of the excitation laser.
		\textbf{b}~Time and fluence dependence of the mean squared displacement, $\Delta\sigma^2(t)$, recorded at $T_N$ along $a$- and $b$-directions (see also Extended Data \cref{fig:ED-Fig.8}~\&~S2).
		Lines indicate linear fits to the data. Images in the inset denote the axis of the transport measurement. 
		\textbf{c}~Position dependence of $X_0$ emission under an excitation fluence of \SI{260}{\micro\J\per\square\cm} (top) and \SI{3100}{\micro\J\per\square\cm} (bottom), corresponding to the estimated exciton densities between $1.1\times 10^{12}$ and $1.3\times 10^{13}$\,cm$^{-2}$ per layer, respectively.
		Sample temperature was nominally 4\,K, but spectral shifts in the region of the laser excitation locally indicate an effective increase in temperature (see Extended Data \cref{fig:ED-Fig.4}). 
		Laser profile is shown by the grey line. 
		\textbf{d}~Schematic illustrating an incoherent magnon flux $\boldsymbol{j}_{mag}$ (blue) propagating away from the excitation region, dragging excitons (red and blue spheres) along. 
		Pulsed laser excitation is indicated by the red line. 
		\textbf{e}~Calculated magnon dispersion. Compared to exciton masses from Ref.~\cite{Klein2022-1}, magnons are substantially heavier than excitons; mass ratios are 38 and 7 along $b$-- and $a$--directions. Solid and dashed lines represent two branches that are very close in energy (see Section S6 for details).
		\label{fig:2}}
\end{figure*}

The photoluminescence (PL) and reflectance contrast spectra of a 9\,nm thin (about 10L, cf. Section S2) crystal in \cref{fig:1}a are typical for the few nanometer-thin CrSBr flakes investigated in our study.
The PL peak at 1.366\,eV matches the well-known resonance of CrSBr excitons ($X_0$) in the reflectance~\cite{Wilson2021,Klein2022-1,Dirnberger2023,Meineke2024} while additional low-energy features in PL are attributed to either phonon-sidebands~\cite{Lin2024} or surface-like states. 
In light of the results presented below, we note that neither the effective diffusion coefficients, nor the emission lifetimes we obtain from our measurements vary significantly across the emission spectrum (see Extended Data \cref{fig:ED-Fig.1} \& Fig.~S14).
The use of such very thin crystals, with purely excitonic optical responses, avoids contributions from self-hybridized polaritons~\cite{Dirnberger2023}.
This allows us to measure the actual propagation of excitons with a transient optical microscopy technique by imaging the spectrally integrated cross-section of the entire PL emission as a function of time and space onto a fast streak camera detector following the excitation by a sub-1\,ps short laser pulse (see Methods)~\cite{Kulig2018}.
Time- and spectrally integrated spatial profiles of the 10L PL signal, presented for 4 and 125\,K in \cref{fig:1}c, already show that the exciton propagation length in CrSBr is temperature dependent.
Even more pronounced are the differences in the time-resolved expansions of the exciton cloud, presented in the inset as a relative increase of the mean squared displacement $\Delta\sigma^2(t)$~\cite{Ginsberg2020}.
	
For a quantitative analysis, we evaluate the effective exciton diffusion coefficient, defined as $D^* = \frac{1}{2} \times \partial \Delta \sigma^2/\partial t$, during the first 80\,ps, as a function of lattice temperature for an excitation fluence of 390\,$\mu$J/cm$^2$, which corresponds to an estimated initial exciton density of about $2\times 10^{12}$\,cm$^{-2}$ per layer (see Extended Data \cref{fig:ED-Fig.2} and Section S3).
As demonstrated in \cref{fig:1}d, the exciton propagation exhibits a pronounced maximum near $T_N$, the critical point of the magnetic phase transition.
Among the observed phenomena characteristic for the spatio-temporal dynamics of excitons in CrSBr (Extended Data \cref{fig:ED-Fig.3}), the temperature dependence of $D^*$ is particularly intriguing because of its striking similarity with the nearly diverging magnetic susceptibility at $T_N$~\cite{Long2023}.
This correlation suggests that the transport of excitons is not determined by classical diffusion or hopping as in the majority of semiconductors. 
Instead, the coupling of excitons to the spin degree of freedom seems to play a major role.

This notion is strongly supported by two key findings of our study. 
First, at $T=T_N$, exciton transport is almost isotropic with respect to the $a$- and $b$-axis, as shown by the symmetric PL shape and similar density-dependent traces of $\Delta\sigma^2$ in \cref{fig:2}a,b.
This observation is in stark contrast with the strongly anisotropic dispersion of excitons and electrons~\cite{Wilson2021,Klein2022-1,Bianchi2023-1,Bianchi2023-2} and the anisotropic electric conductivity~\cite{Wu2022} in CrSBr, but in good agreement with recent studies reporting nearly isotropic magnon transport~\cite{Bae2022,Sun2024}.
Our calculation of the magnon dispersion in \cref{fig:2}e further shows that magnons are not only much more isotropic, but also much heavier than excitons.
Second, the time-resolved expansion of the exciton cloud shown in \cref{fig:2}b strongly depends on excitation fluence and can thus become exceptionally fast.
Most values of $D^*$ we obtain from evaluating the dynamics in the first 20\,ps are orders of magnitude larger than those expected from a classical diffusion model.
The latter estimates exciton diffusion coefficients to be in the 1\,cm$^2$/s range, or below, based on exciton masses in $a$- and $b$-directions and scattering rates obtained from the temperature-dependent linewidths of the $X_0$ peak (cf. Section S5A and orange lines in \cref{fig:1}d). 
	
Besides diffusion, few other processes are known to impact exciton transport.
Among them, exciton-exciton repulsion can be excluded due to its strong dependence on the effective mass and the expected anisotropy~\cite{Vogele2009}.  
Exciton-exciton annihilation~\cite{Kumar2014,Kulig2018} may potentially play a role, as it can lead to an apparent, density-dependent broadening of the spatial exciton distribution.
However, the annihilation coefficients we obtain and the resulting contributions to the effective diffusion coefficient significantly underestimate our observations for the majority of the studied experimental conditions with the exception of room temperature (see Fig.~S9).
Most importantly, one would not expect this process to be enhanced specifically at the Néel temperature.
Due to the general importance of this process for other two-dimensional semiconductors, we provide an extensive discussion of exciton-exciton annihilation in the Section 5B of the Supplementary Information.
	
The position-dependence of the $X_0$ emission peak at $T=4$\,K, where the reduction in linewidth resolves much smaller spectral shifts (see \cref{fig:2}c), reveals that significant amounts of excess energy are released in the CrSBr crystals upon optical excitation.
Especially under higher fluence, sizable spectral shifts within the excitation area strongly indicate local heating by several tens of Kelvin, as shown by the analysis in Extended Data \cref{fig:ED-Fig.4}. 
Thus, not only excitons but also an imbalance in the spatial occupation of phonons and magnons is created by the optical excitation~\cite{Kirilyuk2010}, resulting in a flux of all quasiparticles away from the excitation region~\cite{Au2013,An2016}.
While neither the pure Seebeck drift of excitons themselves~\cite{Perea-Causin2019} nor their coupling to phonons~\cite{Bulatov1992, Glazov2019} can explain the peak we observe at the Néel temperature, thermal magnon currents, as we discuss below, are key for understanding the transport of excitons in this material. 
It is also worth noting that the high sensitivity of the PL to exciton populations in the first tens of picoseconds, the larger fluence, and the absence of external fields contrast the nanosecond propagation dynamics of coherent magnons studied in recent pump-probe experiments~\cite{Bae2022,Diederich2022,Sun2024}.

\begin{figure*}[h!]
	\includegraphics[width=1.0\linewidth]{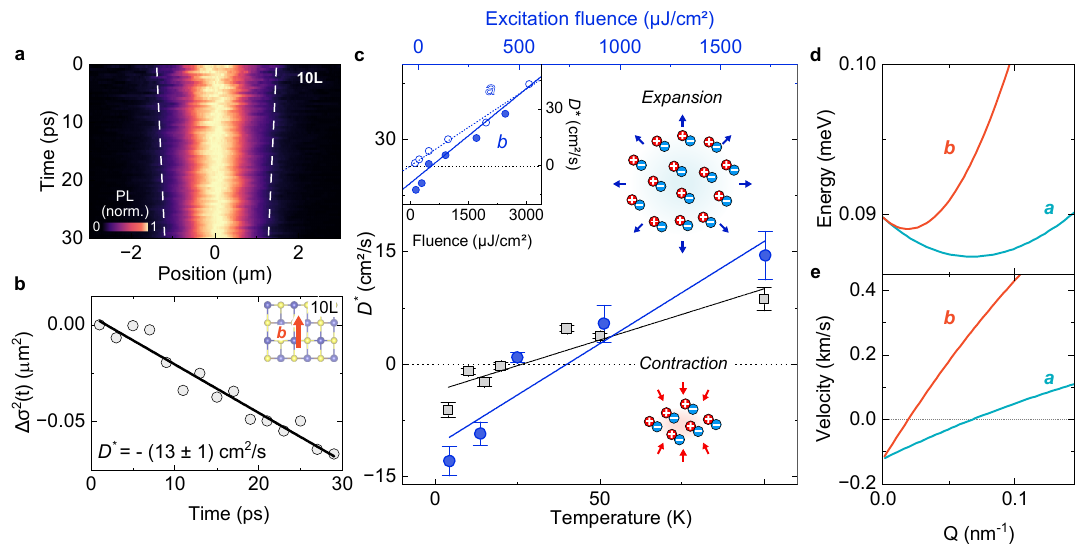}
	\caption{\textbf{Effective contraction of the exciton cloud.} 
		\textbf{a}~Streak camera image from 10L recorded at $T=4$\,K showing space- and time-resolved exciton PL normalized to maximum at each time step. Exciton transport measured along the $b$--direction under a laser excitation fluence of \SI{155}{\micro\J\per\square\cm}. Dashed lines are a guide to the eye. 
		\textbf{b}~Line fit of the mean squared displacement, $\Delta\sigma^2(t)$, extracted from \textbf{a} corresponds to an effective diffusion coefficient of $D^* = -(13\pm1)$\,cm$^2$/s. 
		\textbf{c}~Dependence of $D^*$ on temperature (gray squares) and excitation fluence (blue circles) shows a transition from exciton contraction to expansion. Excitation fluence for the temperature series was \SI{310}{\micro\J\per\square\cm}.
		Inset: Excitation fluence dependence of $D^*$ measured at 4\,K in the 10L crystal for transport along $a$- and $b$-axis. 
		Solid lines approximate a linear fluence dependence. 
		\textbf{d}~Magnon dispersion of the lowest branch for small momenta. 
		\textbf{e}~Corresponding magnon velocities.  
		\label{fig:3}}
\end{figure*}

We propose that a mutual drag between excitons and thermal magnon currents emerges directly from their scattering.
Our theoretical analysis in Section S7 demonstrates that the underlying interaction is distinct from the exciton-magnon coupling recently observed in the canted spin state. 
Because of their larger effective mass and occupation, magnons are able to substantially accelerate excitons through scattering, as indicated in \cref{fig:2}d, analogous to the effects enhancing the thermal transport of electrons in magnetic materials~\cite{Blatt1967}.
This process of magnon-exciton drag qualitatively explains key signatures of our experiments:
First of all, the steady increase in the population of thermal magnons upon approaching the Néel temperature enhances the magnon flux~\cite{Qiu2016,Li2019,Tu2020} and thus maximizes the magnon-exciton drag effect, which is in good agreement with the pronounced maximum of $D^*$ we observe at $T_N$.
This is also confirmed by a recent study on CrSBr~\cite{Canetta2024} reporting a maximum of the electronic Seebeck coefficient near $T_N$.
Even at temperatures above $T_N$, significant drag is expected from the short-range correlations called para-magnons~\cite{Zheng2019} evidenced in magnetometry measurements of CrSBr far beyond $T_N$~\cite{Long2023}.  

A more detailed description of this process is presented in Supplementary Information Section S7.
It estimates that the nearly isotropic dispersion~\cite{Bae2022} and propagation~\cite{Sun2024} of magnons with non-zero momentum may overcome the strong anisotropy of the electronic dispersion when the scattering rates are sufficiently high. 
In this case, the stream of heavy, rapidly propagating magnons essentially carries the excitons along (cf. also \cref{fig:2}d).
This also explains why, at elevated temperatures, anisotropic exciton transport is only observable at very low excitation densities compared to the studied regime (see Extended Data \cref{fig:ED-Fig.5}) and why the differences are much smaller than expected from theory.
Finally, we note that the average expansion of excitons we observe is in overall good agreement with the typical $\sim$km/s propagation velocities of magnons in CrSBr.
Altogether, the magnon-exciton drag effect thus provides a suitable framework for capturing key signatures of the exciton transport observed in our experiments across a broad range of temperatures. 
	
To complete the experimental picture, we now present two particularly striking phenomena observed at low temperatures.
First, at 4\,K, the exciton distribution is not expanding, it appears to be contracting over time, \cref{fig:3}a,b, which can be resolved because the absolute width of the PL spot still exceeds the optical diffraction limit (see Extended Data \cref{fig:ED-Fig.6}). 
The observed contraction is also very fast.
Depending on the chosen model, we either obtain an average inwards velocity of -3\,km/s, or $D^* = -13$\,cm$^2$/s.
Contraction is not too common for excitons~\cite{Ziegler2020,Beret2023} but seems to be ubiquitous in CrSBr, independent of the layer number.
Since the effect is more prominently observed along the $b$--axis, exciton transport at 4\,K is anisotropic in the $ab$--plane. 
Increasing the fluence or the sample temperature, however, turns the anisotropic contraction into a positive, nearly isotropic expansion of the exciton cloud (see \cref{fig:3}c and \ref{fig:2}b). 

The fact that the group velocity of magnons with very small energies and momenta can become negative in CrSBr~\cite{Sun2024} (see \cref{fig:3}d,e) suggests that a scenario in which excitons are dragged by magnons with predominantly antiparallel phase and group velocities is possible. 
A simplified, semi-analytical model presented in section S7 indeed shows that the primary excitation of such magnons at low temperatures and fluences may allow magnons propagating away from the excitation region to scatter excitons backwards, causing a contraction of the exciton cloud.
In contrast, at elevated temperatures or fluences, thermal occupation of magnons with higher energies and momenta, and positive group velocity, favors forward scattering and regular expansion, as observed in \cref{fig:3}c. 
This also motivates a selective excitation of magnons with negative and positive group velocities~\cite{Sun2024} for future experiments in this unusual propagation regime. 

\begin{figure*}[h]
	\includegraphics[width=0.95\linewidth]{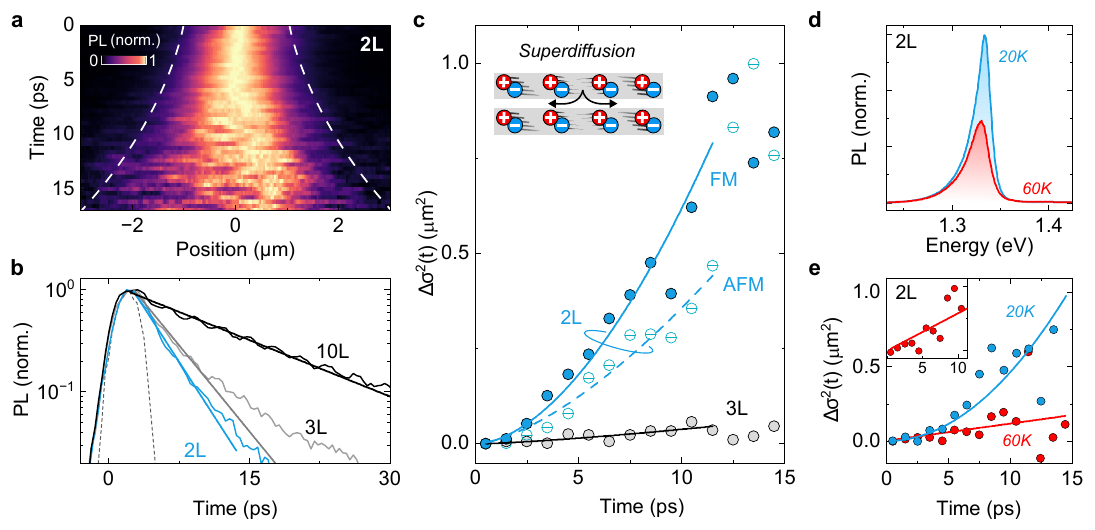}
	\caption{\textbf{Superdiffusive transport in 2L crystals.} 
		\textbf{a}~Streak camera image recorded at $T=4$\,K showing space- and time-resolved exciton PL normalized to maximum at each time step. 
		It shows a rapid expansion in a 2L crystal for FM configuration (see Methods). Excitation fluence is set to \SI{310}{\micro\J\per\square\cm} and the PL is measured along the $b$–axis. Dashed lines are guides to the eye. 
		\textbf{b}~PL transients of the spatially integrated emission. Black dashed line represents the detector response to a $\sim 140$\,fs short excitation pulse. The PL lifetime increases with increasing layer number from 3 ps in 2L (AFM and FM), to 4 ps in 3L, and 11 ps in 10L. 
		\textbf{c}~Corresponding mean squared displacement, $\Delta\sigma^2(t)$, measured for a 2L (AFM and FM) and a 3L (AFM) crystal. The obtained effective propagation velocity of excitons in 2L (FM) is 41\,km/s (see Extended Data \cref{fig:ED-Fig.6}). 
		\textbf{d}~PL spectra recorded at 20\,K and 60\,K. 
		\textbf{e}~$\Delta\sigma^2(t)$ fitted by $\Delta\sigma^2\sim t^\alpha$ with $\alpha>1$ for 20\,K and $\alpha=1$ for 60\,K. Inset shows the 60\,K data in the first 10\,ps.
		\label{fig:4}}
\end{figure*}
	
The second striking observation at low temperature is a remarkably fast expansion of the excitonic emission in 2L crystals spanning hundreds of nanometers within picoseconds (see \cref{fig:4}a).
The broadening is continuous and well-resolved in the first 15\,ps after the excitation and appears to be only limited by the fast decay of excitons shown in \cref{fig:4}b.  
Most interestingly, in this time window, the mean squared displacement $\Delta\sigma^2(t)$ does not increase linearly but exhibits a superlinear behavior.
The observed expansion law, $\Delta \sigma^2 \propto t^\alpha$, with values of $\alpha$ between $1.3\pm0.1$ and $2.1\pm0.3$ obtained from fits, is a hallmark of exciton superdiffusion~\cite{Ginsberg2020}.  
Similar features are consistently observed along the $a$--direction, for both the AFM and the FM phase, and in other 2L samples (see \cref{fig:4}c, Extended Data \cref{fig:ED-Fig.7} \& Methods), yet are absent in a nearby 3L crystal. 
For completeness, we note that monolayer PL signals were too small to draw reliable conclusions.

Superdiffusion generally indicates coherent transport.
However, the ballistic motion of excitons themselves seems an unlikely explanation, since excitons are expected to be frequently scattered by phonons and magnons on these timescales.  
Besides, experimental transport signatures in 2L crystals otherwise match the contraction, nearly isotropic propagation, and pronounced fluence dependence of thicker samples (see Extended Data \cref{fig:ED-Fig.10}).
We thus speculate that the superdiffusive behavior results from enhanced interactions of excitons with ballistically propagating magnon waves. 
This is supported by the fact that the expected transition towards regular diffusion ($\Delta \sigma^2 \propto t$) is observed at temperatures around 60\,K, as shown in Figs.~4e~\&~S1.
The origin of superdiffusion in 2L and the large effective velocities that exceed the velocity of long-range magnon transport reported for bulk CrSBr~\cite{Bae2022,Sun2024} is not clear at this stage. 
Nevertheless, we note that the properties of excitons and magnons in ultrathin crystals could differ from those of bulk (cf. also Figs. S6~\&~7). 
Particularly phenomena related to surface~\cite{Chen2017,Ye2022} and superluminal-like effects~\cite{Lee2021}, as well as a stronger role of phonons~\cite{Liu2024,Bae2024}, could contribute to the ultrafast dynamics of excitons and magnons in 2L crystals.
	
In conclusion, exciton transport in ultrathin crystals of the layered antiferromagnet CrSBr is very fast, fluence dependent, and peaks at the Néel temperature.
It features both expansion and contraction and can become superdiffusive in bilayer crystals.
While common transport mechanisms fail to describe these findings, the scattering of excitons by a flux of thermal magnons is proposed to drive exciton transport.
For sufficiently strong interactions, excitons no longer move independently inside a stream of heavy magnons; they are effectively carried by the magnon current. 
The fact that magnons can exhibit much longer coherence times and lengths than excitons, and may be excited electrically, highlights the considerable potential of exciton-spin interactions to imprint magnon transport properties onto the typically slow motion of excitons.
It might further be possible to drive excitons even in non-magnetic semiconductors by both coherent and incoherent magnon currents using proximity effects in heterostructures.
Altogether, these results are highly promising for the realization of efficient magnetic control of optical quasiparticles, an encouraging new direction for fundamental research on correlated exciton-spin systems and, more broadly, energy and information transport in solids.

	
%
%

\clearpage

\nocite{Telford2022,Smolenski2024,Li2023,Chernikov:2023ab,Glazov:2022aa,Wagner2021,Mouri2014,Sun2014,Yuan2017,Wietek2024,PhysRev.101.944,gantmakher87,ll2_eng,glazov2018electron,Scheie:2022aa,ll9_eng,Akhiezer:1968aa,Gurevich2020,Keffer:1966aa,keldysh_wind,kozub_drag,1996JETP...82.1180K,PSSB:PSSB45}


\section{Methods}
\subsection{Crystal growth and sample fabrication}
CrSBr bulk single crystals were synthesized by the chemical vapor transport method described in Ref.~\cite{Klein2022-1}.
From these bulk crystals, thin flakes with typical lateral extensions of several tens of microns were mechanically exfoliated directly from tape onto standard SiO$_2$/Si substrates with a SiO$_2$ thickness of 285\,nm. 
After the transfer, samples were stored under vacuum conditions.
For the experiments, they were mounted either directly onto the cold finger of a continuous-flow He cryostat, or on top of a small disk magnet with in-plane magnetization providing a permanent magnetic field of $\sim$0.2\,T which was then glued onto the cold finger. 
We estimate an accuracy of $\pm10^\circ$ for the alignment of CrSBr crystals with respect to the in-plane magnetization axis of the magnet and an accuracy of $\pm5^\circ$ for their alignment relative to the detector slit. 
Due to a reduction in the saturation field, 2L crystals placed on top of the disk magnet allow us to study exciton transport in the FM phase inside our cryostat (cf. results presented in \cref{fig:3}).

\subsection{Optical spectroscopy and time-resolved microscopy}
Few-layer crystals with lateral extensions of at least several microns were preselected by optical microscopy.
Their layer number was determined by optical contrast and confirmed by atomic force microscopy. 
Prior to measuring exciton dynamics, each flake was characterized by reflectance and PL spectroscopy.
For reflectance we used the attenuated output of a spectrally broadband tungsten-halogen lamp, focused to a spot size of about $\SI{2.0}{\um}$ by a $\SI{60}{\times}$ glass-corrected microscope objective (NA=0.7).
Spectra measured on top of the bare SiO$_2$/Si substrate were used as a reference for the CrSBr reflectance spectra and analyzed by the transfer-matrix method calculating the absorption spectrum based on a small set of Lorentz oscillators. 

For transient PL microscopy, we used ultrashort ($\sim140$\,fs), linearly polarized optical pulses from a Ti:Sapphire laser tuned to a photon energy of 1.61\,eV, or to 1.77\,eV where specified. 
The laser was focused to a spot size of $\SI{0.8}{\mu m}$ by a $\SI{60}{\times}$ {glass-corrected} objective.
For each flake, the linear polarization of the laser was aligned parallel to the crystallographic $b$--axis; no polarization-selective optics were used for the detection of the emission. 
The PL was spectrally filtered to remove the laser excitation before being directed into the spectrometer where it was either spectrally dispersed by a grating or imaged by a silver mirror. 
The signal was detected by a charge-coupled device and by a streak camera to acquire time-integrated and time-resolved data, respectively.
We estimate the accuracy of $D^*$ values determined by our experiment to be $\pm$0.1\,cm$^2$/s. 

In our experiments, the variance $\sigma$ of the PL spot in the first one or two picoseconds is typically $\sim0.3\mu m$ larger than the size of the laser spot. 
This could result from differences in the excitation and detection wavelengths, chromatic aberration of the imaging system, as well as potential ultrafast, sub-picosecond propagation processes that are beyond the resolution of the streak camera detector. 

\noindent\textbf{Acknowledgments:}
Financial support by the DFG via SFB 1277 (project B05, Project-ID: 314695032), the Würzburg-Dresden Cluster of Excellence on Complexity and Topology in Quantum Matter (ct.qmat) (EXC 2147, Project-ID 390858490), and the Emmy Noether Program (F.D., Project-ID 534078167) is gratefully acknowledged. 
A.C. acknowledges funding from ERC through CoG CoulENGINE (GA number 101001764). 
Z.S. was supported by ERC-CZ program (project LL2101) from Ministry of Education Youth and Sports (MEYS) and used large infrastructure from MEYS project reg. No. CZ.02.1.01/0.0/0.0/15\_003/0000444 financed by the ERDF.
A.K. acknowledges financial support from the Spanish Ministry for Science and Innovation -- AEI Grant CEX2018-000805-M (through the "Maria de Maeztu" Programme for Units of Excellence in R\&D) and grant RYC2021-031063-I funded by MCIN/AEI/10.13039/501100011033 and "European Union Next Generation EU/PRTR".\\

\noindent\textbf{Author contributions:}
F.D. and A.C. conceived the experimental idea. 
F.D. and S.T. performed the experiments.
S.T. fabricated the samples and F.D. analyzed the data.
K.M. and Z.S. provided the bulk crystals.
M.M.G., A.K., and Z.A.I. provided theoretical support.
The manuscript was written by F.D. and A.C. with input from all authors.\\

\noindent\textbf{Competing interests:} 
The authors declare that they have no competing interests.\\

\noindent\textbf{Data availability:} 
The main data sets generated and analyzed in this study are available at [insert link in proof stage]. All other data will be provided by the corresponding authors upon request.\\


\newpage
\renewcommand{\theequation}{\arabic{equation}}%
\renewcommand{\figurename}{EXTENDED DATA FIG.}
\renewcommand{\thesection}{\arabic{section}}
\newcounter{ffigure}
\renewcommand{\thefigure}{\arabic{ffigure}}
\setcounter{ffigure}{1}

\begin{figure*}[h!] 
	\includegraphics[scale=0.8]{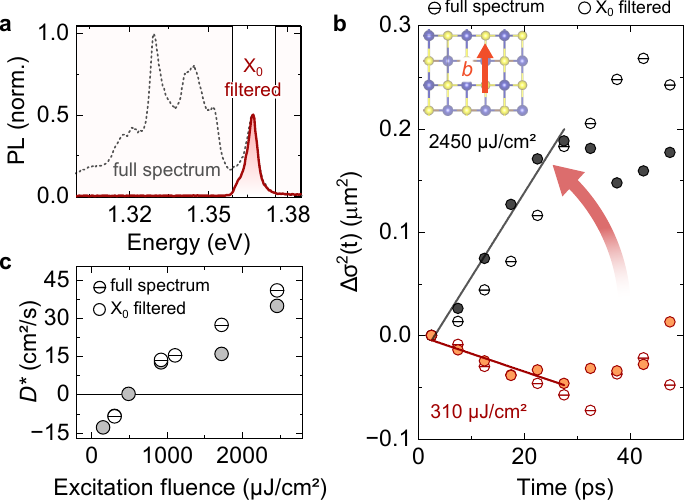}
	\caption{\textbf{Exciton transport at 4K measured in 10L for the full spectrum (unfiltered) and for the $X_0$ peak (filtered).}
		\textbf{a}~Spectrally filtered (red) PL spectrum of the 10L flake compared to the unfiltered (grey) spectrum. Shaded areas indicate the spectral cut-off of the filters. Both spectra are normalized to the maximum intensity of the full spectrum. 
		\textbf{b}~Variation of the mean squared displacement, $\Delta \sigma^2(t) = \sigma^2(t) - \sigma^2(0)$, obtained along the $b$--axis for the full spectrum and the filtered spectrum under 310 and 2450 \SI{}{\micro\J\per\square\cm} excitation fluence.
		\textbf{c}~Excitation fluence dependence of $D^*$ for both cases determined by evaluating the first 30\,ps.  
		\label{fig:ED-Fig.1}}\addtocounter{ffigure}{1}
\end{figure*}

\newpage

\begin{figure*}[h!]
	\includegraphics[scale=0.8]{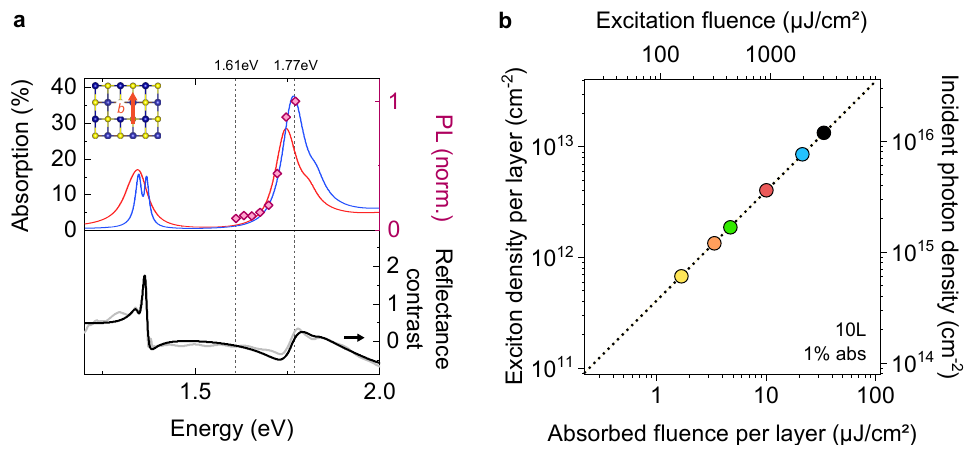}
	\caption{\textbf{Calculated optical absorption and exciton density in the 10L crystal.} 
		\textbf{a}~Reflectance contrast spectrum measured at 4\,K is fitted by a Lorentz oscillator model to calculate the absorption spectrum. Dashed line indicates the excitation energy of the optical pulses used to study exciton transport. 
		\textbf{b}~Excitation fluence and exciton and photon density for 1\% absorption. Color coding of circles matches that of Fig. 2 and others.
		\label{fig:ED-Fig.2}}\addtocounter{ffigure}{1}
\end{figure*}

\newpage

\begin{figure*}[h!]
	\includegraphics[scale=0.8]{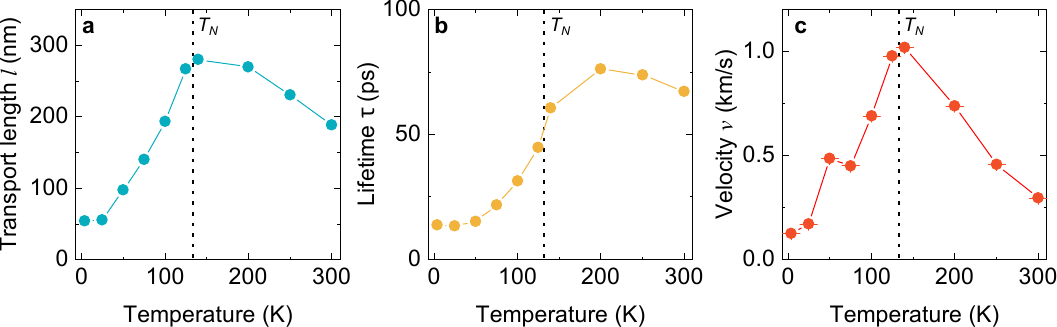}
	\caption{\textbf{Temperature dependence of spatio-temporal exciton dynamics.} 
		\textbf{a}~Exciton transport length $l$ determined by $l=\sqrt{2D^*\tau}$.
		\textbf{b}~Exciton lifetime $\tau$ determined from single exponential fits of the spatially integrated exciton decay measured for different temperatures (cf. discussion in Section S9).
		\textbf{c}~Effective exciton propagation velocity, $v=\Delta\sigma(t)/t$, evaluated from a linear fit of $\Delta\sigma(t)$ over the first 20\,ps after excitation (cf. also \cref{fig:ED-Fig.6}). 
		All data obtained under an excitation fluence of 390\,$\mu$J/cm$^2$. 
		\label{fig:ED-Fig.3}}\addtocounter{ffigure}{1}
\end{figure*}

\newpage

\begin{figure*}[h!]
	\includegraphics[scale=0.8]{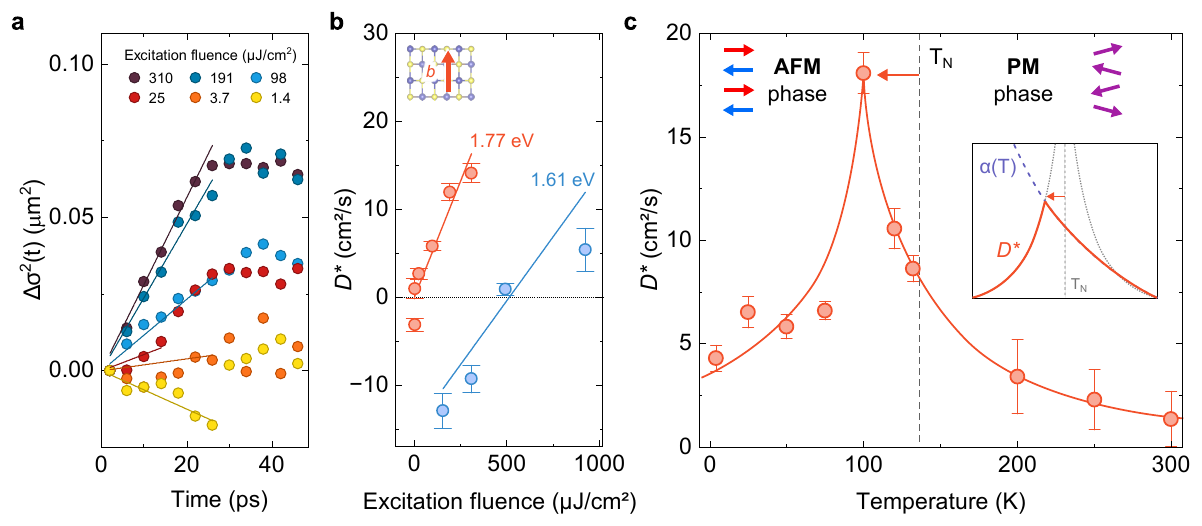}
	\caption{\textbf{Exciton transport dynamics in 10L measured under resonant excitation of the B-exciton at 1.77\,eV.} 
		\textbf{a}~Fluence dependence of the mean squared displacement, $\Delta\sigma^2(t)$, recorded at T = 4\,K.  
		\textbf{b}~Fluence dependence of the effective diffusion coefficient obtained under 1.77\,eV (red) and 1.61\,eV (blue) excitation energies.  
		\textbf{c}~Temperature dependence of the effective diffusion coefficient measured with a fluence of \SI{20}{\micro\J\per\square\cm} for the excitation photon energy of 1.77\,eV. Solid red line is a guide to the eye. 
		The inset illustrates how a temperature-dependent decrease of the absorption coefficient, $\alpha(T)$, indicated by the blue dashed line, can shift the maximum of $D^*$ towards lower temperatures.   
		\label{fig:Fig-R2}}\addtocounter{ffigure}{1}
\end{figure*}	

\newpage

\begin{figure*}[h!]
	\includegraphics[scale=0.8]{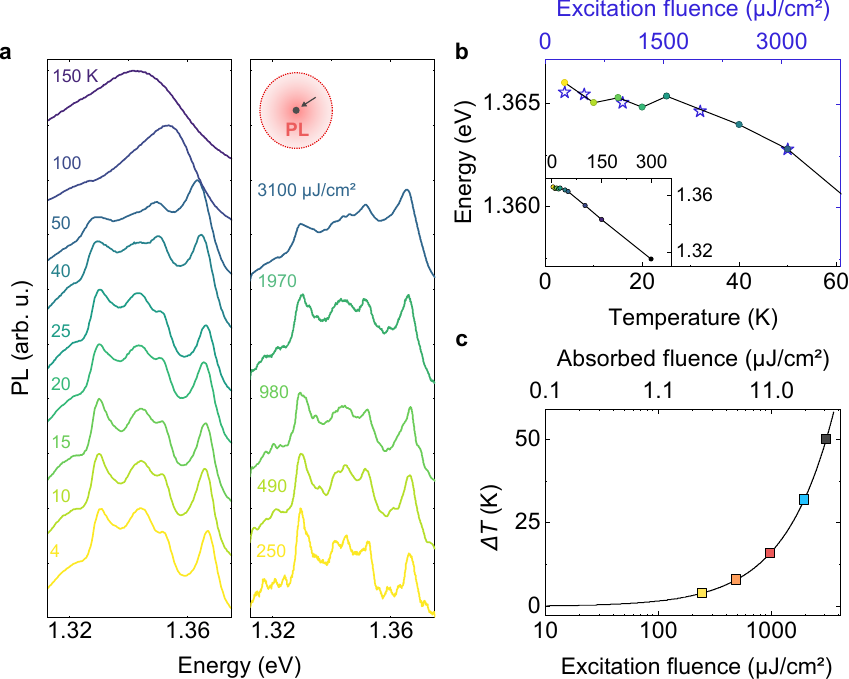}
	\caption{\textbf{Estimated increase of the effective sample temperature under pulsed optical excitation.} 
		\textbf{a}~Left: PL emission of 10L as a function of temperature recorded under 310\,$\mu$J/cm$^2$ fluence. Right: Same PL but as a function of excitation fluence at 4\,K. Red circle and black marked spot indicate the full size of the PL emission spot and the center of the diffraction-limited PL spectra taken for data analysis.
		\textbf{b}~Overlaid temperature and fluence dependence of the $X_0$ emission energy at nominally 4\,K determined in \textbf{a}. Inset shows a larger range of the temperature dependence. 
		\textbf{c}~Interpolated excitation-induced linear increase in temperature $\Delta T$.  
		\label{fig:ED-Fig.4}}\addtocounter{ffigure}{1}
\end{figure*}

\newpage

\begin{figure*}[h!]
	\includegraphics[scale=0.8]{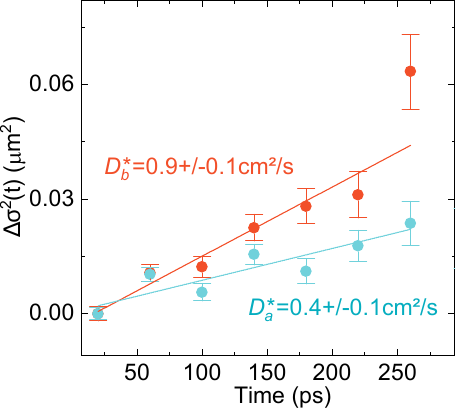}
	\caption{\textbf{Signatures of anisotropic exciton transport.}
		Variation of the mean squared displacement, $\Delta\sigma^2(t)$, measured along $a$ and $b$ in 10L at 300\,K for a small fluence, compared to the studied regime throughout the main manuscript, of 55~\SI{}{\micro\J\per\square\cm} corresponding to exciton density of about $2\times 10^{11}$\,cm$^{-2}$ per layer. 
		\label{fig:ED-Fig.5}}\addtocounter{ffigure}{1}
\end{figure*}

\newpage

\begin{figure*}[h!]
	\includegraphics[scale=0.8]{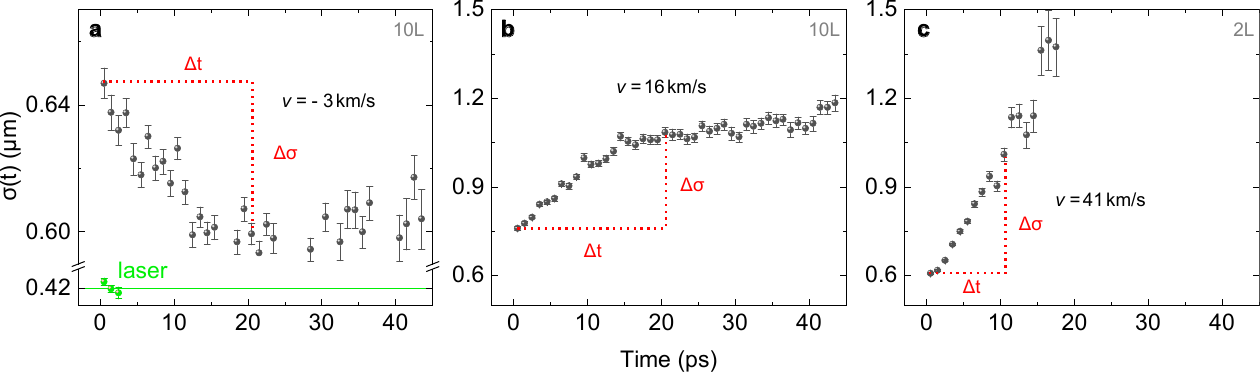}
	\caption{\textbf{Effective exciton velocity determined from transport measurements.} 
		Exemplary measurement of $\sigma(t)$ for \textbf{a}~10L and 310\,$\mu$J/cm$^2$, \textbf{b}~10L and 3100\,$\mu$J/cm$^2$, \textbf{c}~2L and 310\,$\mu$J/cm$^2$ illustrating the estimation of exciton propagation velocities as $v = \Delta \sigma/\Delta t$. Note that similar values can be found from a linear fit of $\sigma(t)$ for $t \lesssim 15-20$\,ps. The values of $\sigma$ obtained when imaging the laser (green dots) directly onto the streak camera do not change with time and are close to 0.4\,$\mu$m. All data recorded at $T=4$\,K.
		\label{fig:ED-Fig.6}}\addtocounter{ffigure}{1}
\end{figure*}

\newpage

\begin{figure*}[h!]
	\includegraphics[scale=0.7]{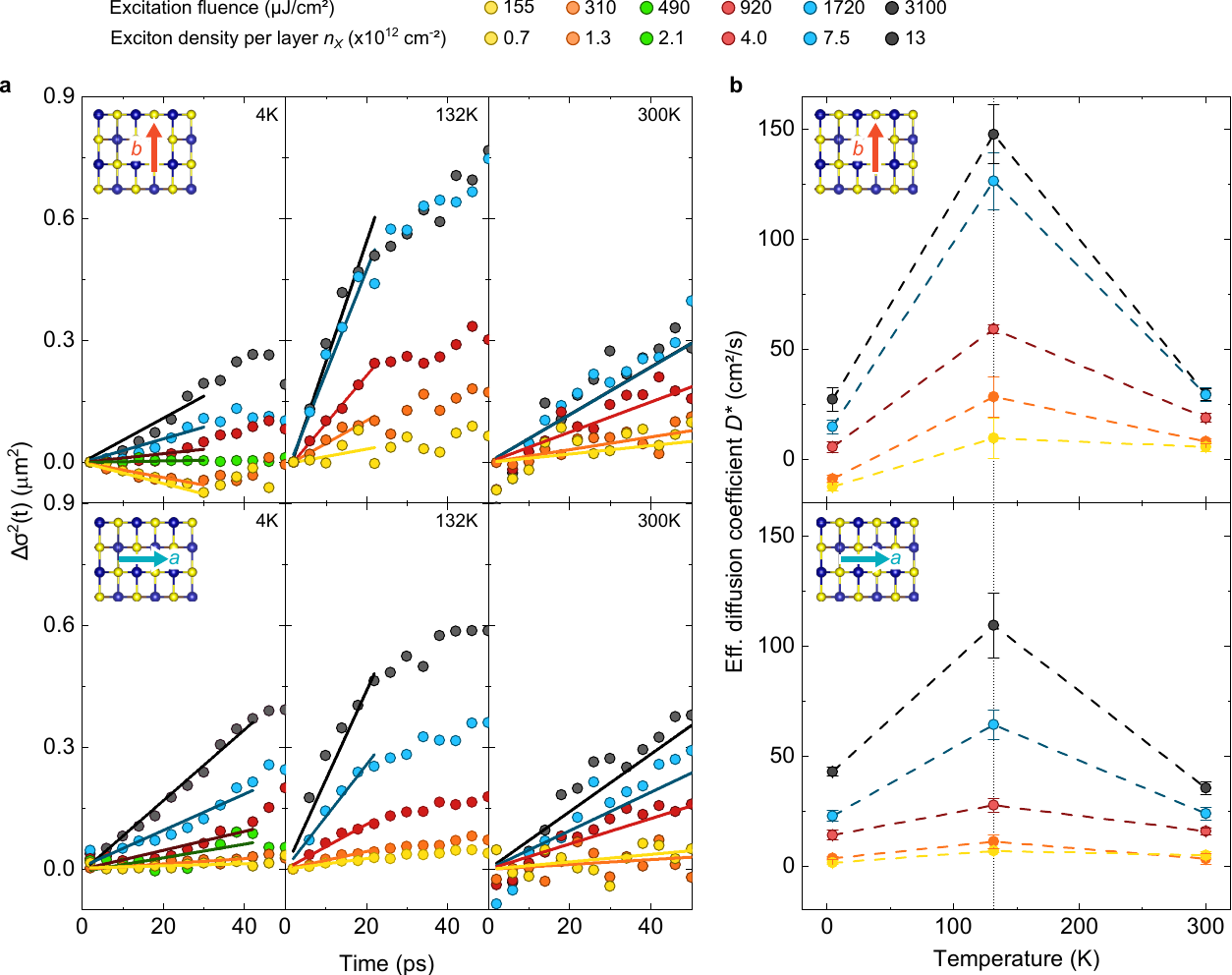}
	\caption{\textbf{Exciton transport in 10L along \textit{a} and \textit{b} at 4\,K, 132\,K, and 300\,K.} 
		\textbf{a}~Mean squared displacement, $\Delta\sigma^2(t)$, measured for different excitation fluence. Solid lines indicate the linear fits to obtain the values $D^*$ shown in Fig.~S9. 
		\textbf{b}~Effective diffusion coefficients $D^*$ obtained from the data in \textbf{a}. 
		\label{fig:ED-Fig.8}}\addtocounter{ffigure}{1}
\end{figure*}

\newpage

\begin{figure*}[h!]
	\includegraphics[scale=0.8]{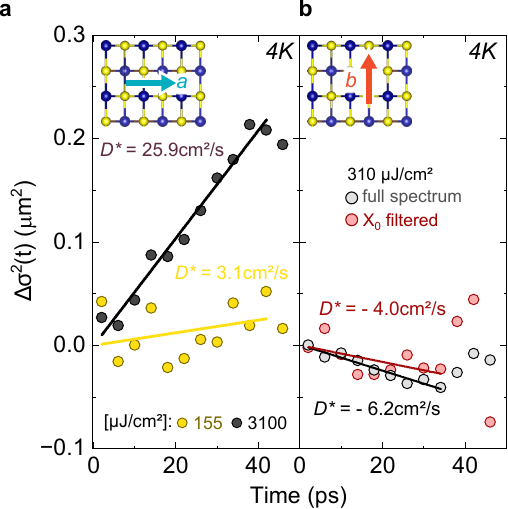}
	\caption{\textbf{Exciton transport at 4\,K in a 11L crystal.} 
		\textbf{a}~Exciton transport measured along the $a$--direction for 155 (yellow) and 3100 (black) \SI{}{\micro\J\per\square\cm}. 
		\textbf{b}~Exciton transport measured along $b$ for the full spectrum (unfiltered) and for the $X_0$ peak (filtered). Both measurements were recorded with an excitation fluence of 310 \SI{}{\micro\J\per\square\cm} and evaluated during the first 35\,ps.
		\label{fig:ED-Fig.10}}\addtocounter{ffigure}{1}
\end{figure*}	

\newpage

\begin{figure*}[h!]
	\includegraphics[scale=0.8]{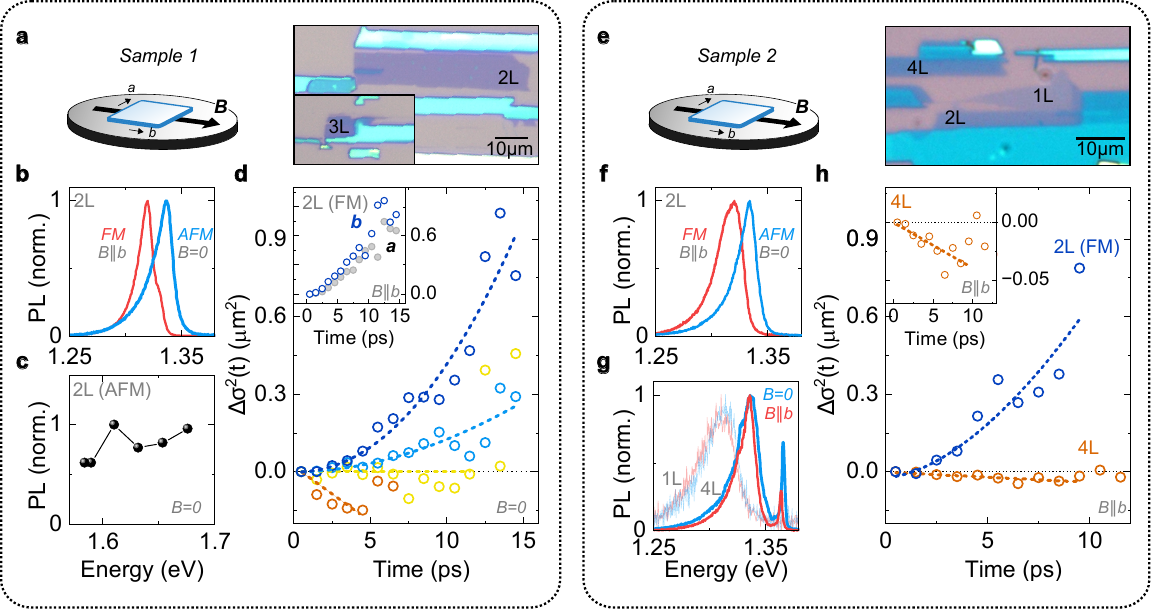}
	\caption{\textbf{Exciton transport in 2L and other few-layer crystals at 4\,K.} 
		\textbf{a}~Left: Schematic of the sample chip mounted on top of a small disk magnet with in-plane magnetization. Right: Optical microscope image of a 2L and a 3L crystal on the chip, which is glued onto the magnet such that the $b$--axis of the crystals aligns with the magnetization axis with an estimated precision of $\pm10^\circ$. 
		\textbf{b}~PL emission of the 2L crystal when the chip is mounted together with the magnet (red), or directly on top of the cold finger of the cryostat (blue). The spectral shift of the PL indicates a field-induced transition of the magnetic order into an FM state~\cite{Wilson2021}. 
		\textbf{c}~Integrated PL signal of the 2L crystal ($B=0$, AFM) shows only a weak dependence on excitation energy. 
		\textbf{d}~Fluence dependence of $\Delta\sigma^2(t)$ measured along the $b$--axis of the 2L crystal for 30 (orange), 55 (yellow), 150 (light blue), 310 (dark blue) \SI{}{\micro\J\per\square\cm} in the AFM phase without the magnet at B=0.
		Dashed lines are guides to the eye. 
		Inset: $\Delta\sigma^2(t)$ measured along $a$ and $b$--axis in the FM phase on top of the magnet ($B\parallel b$) with 310 \SI{}{\micro\J\per\square\cm}. 
		\textbf{e,f}~Analogous to \textbf{a},\textbf{b} on a second sample.
		\textbf{g}~PL emission of a 1L and a 4L crystal when the chip is mounted together with the magnet (red), or directly on top of the cold finger of the cryostat (blue), illustrating the lack of energy shifts, as expected~\cite{Ye2022}. 
		\textbf{h}~Measurement of $\Delta\sigma^2(t)$ along the $b$--axis of a 2L (blue) and a 4L (orange) crystal for 500 \SI{}{\micro\J\per\square\cm} on top of the magnet. The magnetic configuration is FM for the 2L crystal but because of the larger switching field required remains AFM for the 4L crystal. Dashed lines are guides to the eye. Inset: Magnified view of the negative transport measured in the 4L crystal. 
		All data recorded at 4\,K.
		\label{fig:ED-Fig.7}}\addtocounter{ffigure}{1}
\end{figure*}

\end{document}


\title{SUPPLEMENTARY INFORMATION:\\Exciton transport driven by spin excitations in an antiferromagnet} 

\author{Florian Dirnberger$^\ddagger$}
\email{f.dirnberger@tum.de}
\affiliation{Institute of Applied Physics and Würzburg-Dresden Cluster of Excellence ct.qmat, TUD Dresden University of Technology, Dresden, 01187, Germany}
\affiliation{Zentrum für QuantumEngineering (ZQE), Technical University of Munich, Garching, Germany}
\affiliation{Department of Physics, TUM School of Natural Sciences, Technical University of Munich, Munich, Germany}
\thanks{Authors contributed equally.}	

\author{Sophia Terres$^\ddagger$}
\affiliation{Institute of Applied Physics and Würzburg-Dresden Cluster of Excellence ct.qmat, TUD Dresden University of Technology, Dresden, 01187, Germany}
\thanks{Authors contributed equally.}	

\author{Zakhar A. Iakovlev}
\affiliation{Ioffe Institute, 194021 St. Petersburg, Russia}

\author{Kseniia Mosina}
\author{Zdenek Sofer}
\affiliation{Department of Inorganic Chemistry, University of Chemistry and Technology Prague, Technickaá 5, 166 28 Prague 6, Czech Republic}

\author{Akashdeep Kamra}
\affiliation{Department of Physics and Research Center OPTIMAS, Rheinland-Pf\"alzische Technische Universit\"at Kaiserslautern-Landau, 67663 Kaiserslautern, Germany}
\affiliation{Departamento de Física Teórica de la Materia Condensada and Condensed Matter Physics Center (IFIMAC), Universidad Autónoma de Madrid, E- 28049 Madrid, Spain}

\author{Mikhail M. Glazov}
\affiliation{Ioffe Institute, 194021 St. Petersburg, Russia}

\author{Alexey Chernikov}
\email{alexey.chernikov@tu-dresden.de}
\affiliation{Institute of Applied Physics and Würzburg-Dresden Cluster of Excellence ct.qmat, TUD Dresden University of Technology, Dresden, 01187, Germany}
\maketitle
\tableofcontents
\clearpage

\nocite{Seyler2018,Zhang2019,Kang2020,Wilson2021,Baibich1988,Binasch1989,Bae2022,Diederich2022,Dirnberger2023,Unuchek2018,Dong2021,Jakhangirkhodja2023,Birowska2021,Kim2022,Barman2021,Lebrun2018,Tu2020,Lee2021,Hortensius2021,Wei2022,Sun2024,Goser1990,Telford2020,Long2023,Canetta2024,Klein2022-1,Tabataba2023,Meineke2024,Lin2024,Kulig2018,Ginsberg2020,Bianchi2023-1,Bianchi2023-2,Wu2022,Vogele2009,Kumar2014,Kirilyuk2010,Au2013,An2016,Perea-Causin2019,Bulatov1992,Glazov2019,Blatt1967,Qiu2016,Li2019,Zheng2019,Ziegler2020,Beret2023,Chen2017,Ye2022,Liu2024,Bae2024}

\section{Supplementary figures}

\begin{figure*}[h!]
	\includegraphics[scale=0.8]{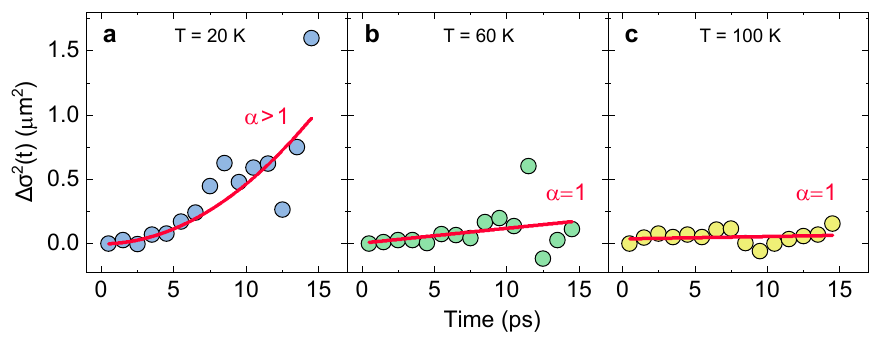}
	\caption{\textbf{Thermal breakdown of exciton superdiffusion in 2L.} 
	Exciton transport in 2L measured at \textbf{a} T = 20K, \textbf{b} 60K, and \textbf{c} 100K. Excitation fluence was 310 \SI{}{\micro\J\per\square\cm} and the transport was measured in the $b$--direction. 
	\label{fig:Fig-R1}}
\end{figure*}	

\newpage




\begin{figure*}[h!]
	\includegraphics[scale=0.8]{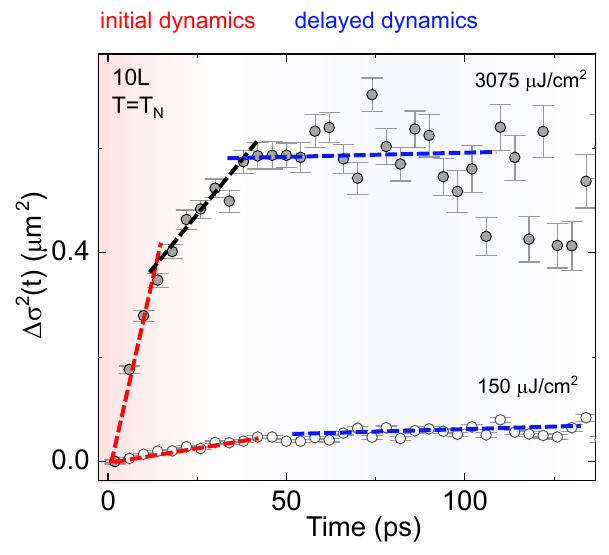}
	\caption{\textbf{Exciton transport in 10L at $T=T_N$.}
	Comparison between the spatiotemporal dynamics of excitons observed at low and high excitation fluence along the $b$--axis. 
	Particularly at large fluence, two distinct slopes of $\Delta\sigma(t)$ indicate distinct transport regimes labeled initial and delayed dynamics are noticeable. 
	\label{fig:Fig-R9}}
\end{figure*}	

\clearpage
\pagebreak

\section{Sample details}

\begin{figure*}[h!]
	\includegraphics[scale=1]{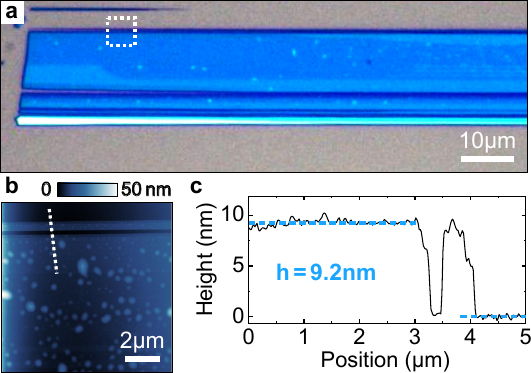}
	\caption{\textbf{Determination of the crystal thickness.} 
		\textbf{a}~Optical microscope image of the 10L crystal. 
		\textbf{b}~Atomic force microscopy image recorded at the position marked by the white square in \textbf{a}. 
		\textbf{c}~Height profile obtained at the position indicated by the white dashed line determines a crystal thickness of $h=9.2$\,nm. Following Ref.~\cite{Telford2022}	and the empirical relation $h=0.79N+1.12$\,nm, this thickness corresponds to a layer number of $N=10$.
		\label{fig:SI-Fig-0}}
\end{figure*}

\clearpage
\pagebreak

\section{Spectral analysis of PL emission spectra under increasing laser excitation fluences}

In this section we provide arguments and analysis regarding estimated photo-excited electron-hole-pair densities and compare them to the Mott density where excitons dissociate into dense electron-hole plasma. 
The analysis presented below suggest that excitons dominate the photo-excited density of optical quasiparticles even at the largest excitation fluences applied in our study. 
Strong confinement of excitons by the magnetic AFM configuration in the out-of-plane direction, as well as the large effective exciton mass ($M_a=10.2m_0$) along the in-plane $a$--direction lead to excitons with very small radii and a quasi-1D density of electronic states~\cite{Smolenski2024}.
Both render the excitons in CrSBr particularly robust against screening by free electron-hole pairs in contrast to other systems, such as semiconducting transition-metal dichalcogenides, for example.

\paragraph*{Theoretical considerations.} In our 10L crystal, under 1.61eV excitation, the largest exciton density we estimate per layer is $1.3\times{10}^{13}\ {cm}^{-2}$. Using the exciton Bohr radii ($a_B^b=$4.2Å along b- and $a_B^a=$1.3Å along $a$--direction) reported in a recent study~\cite{Li2023}, the approximate relation $n_M\approx\ 1/(\pi a_B^a a_B^b)$ leads to a Mott density of $n_M\approx\ 6\times{10}^{14}\ {cm}^{-2}$.
Thus, while this relation is known to slightly overestimate the Mott density, even for the largest excitation fluences used our study, exciton densities are still more than one order of magnitude below the Mott transition. 
For the key results presented in Fig. 1, for example, the estimated densities were substantially smaller, one the order of ${10}^{12}\ {cm}^{-2}$ per layer. 

\begin{figure*}[h!]
	\includegraphics[scale=0.8]{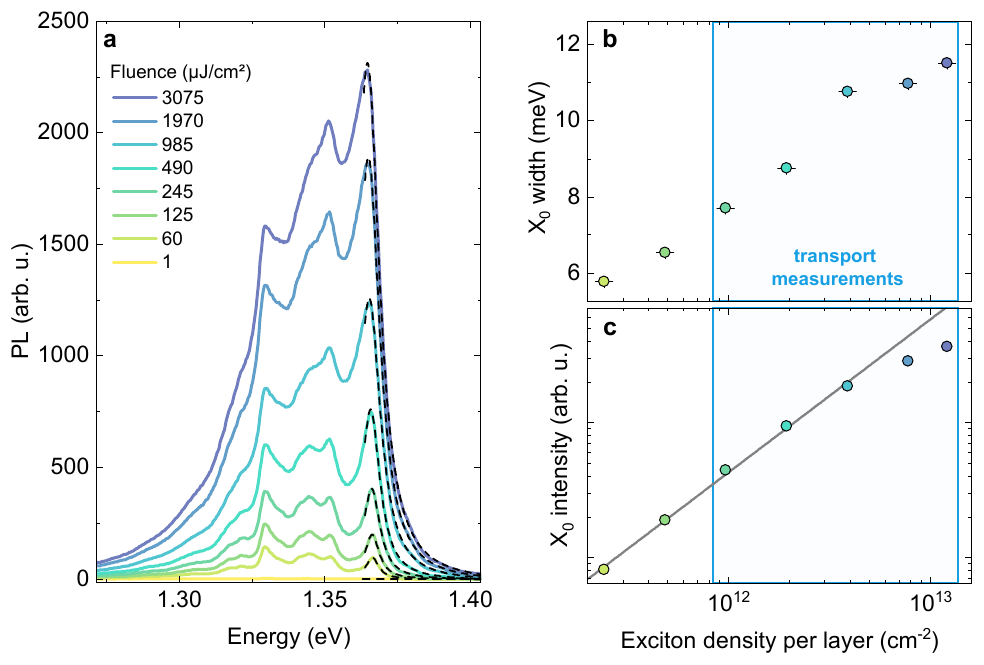}
	\caption{\textbf{Spectral analysis of the fluence-dependent PL emission at T = 4\,K in the 10L crystal.} 
		\textbf{a}~PL emission spectra obtained under different excitation fluences. 
		\textbf{b}~Spectral width of the $X_0$ exciton emission peak determined from Lorentzian fits (black dashed lines in \textbf{a}) as a function of exciton density.  
		\textbf{c}~Corresponding integrated intensity of the $X_0$ emission peak. 
		Conversion of excitation fluence to exciton density as shown in Extended Data Fig. 2. Blue-shaded rectangles mark the regime used to study exciton transport. Laser excitation energy was set to 1.61\,eV. 
		\label{fig:SI-Fig-R4}}
\end{figure*}

\paragraph*{Fluence dependence in a 10L crystal.} Overall, this theoretical estimate seems to be in good agreement with the fluence-dependent PL spectra measured in the 10L crystal. While we observe a broadening of the PL emission spectrum, as expected for increasing fluence, the absolute increase in linewidth is rather moderate. The PL data presented in Fig. R1 shows that the $X_0$ linewidth increases from 6\,meV at the lowest fluence to 11\,meV at the highest excitation fluence applied in our study. Even at this fluence, the $X_0$ exciton peak is still clearly visible and does not change its spectral shape. Moreover, we find that the integrated PL intensity of the $X_0$ peak increases linearly up to exciton densities as large as $\sim5\times{10}^{12}\ {cm}^{-2}$. 
As also noted above, most of the transport measurements reported in our study are conducted in this linear regime, suggesting the optical excitations are predominantly comprised of excitons. Even at larger densities, the main effect appears to be the onset of the saturation of the PL emission signal, not exciton dissociation.

\begin{figure*}[h!]
	\includegraphics[scale=0.8]{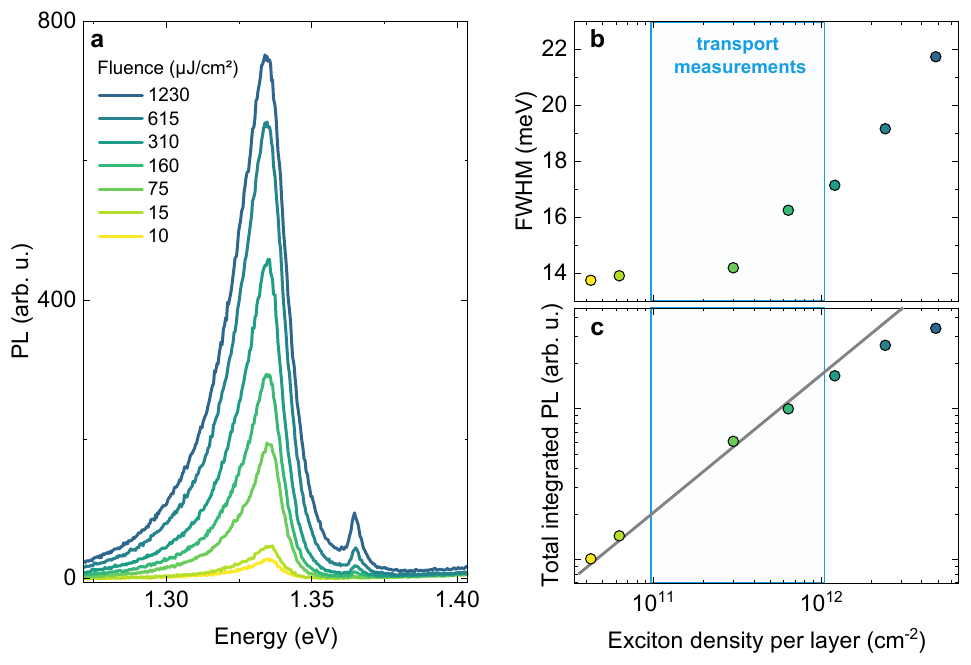}
	\caption{\textbf{Spectral analysis of the fluence-dependent PL emission in a 2L crystal at T = 4\,K.} 
		\textbf{a}~PL emission spectra obtained under different excitation fluences. 
		\textbf{b}~Full width at half maximum (FWHM) of the PL signal determined from the numerical integral as a function of the exciton density.
		\textbf{c}~Corresponding integrated intensity of the full PL spectrum under increasing exciton densities. Blue-shaded rectangles mark the regime used to conduct the majority of 2L transport measurements. Laser excitation energy was set to 1.61\,eV.
		\label{fig:SI-Fig-R3}}
\end{figure*}

\paragraph*{Fluence dependence in a 2L crystal.} It is important to note that much smaller fluences are required in order to observe pronounced transport phenomena in 2L, because exciton transport in these crystals is superdiffusive and exceptionally fast. Hence, our 2L transport measurements (with very few exceptions) were conducted entirely in the linear response regime. This regime is marked by the gray rectangle in Fig. R2. It is characterized by a linear scaling of the integrated PL intensity as a function of excitation fluence, consistent with the excitonic regime below the Mott transition. Moreover, we note that the linewidth of the 2L emission increases only slightly, from 14\,meV to 17\,meV, similar to the thicker samples. Strong screening effects that dissociate the excitons into an electron-hole plasma are therefore not expected at the fluences we used to study exciton transport also in the 2L crystals. 	

\clearpage
\pagebreak

\section{Differences in the optical properties of 2L and 10L}

\begin{figure*}[h!]
	\includegraphics[scale=0.8]{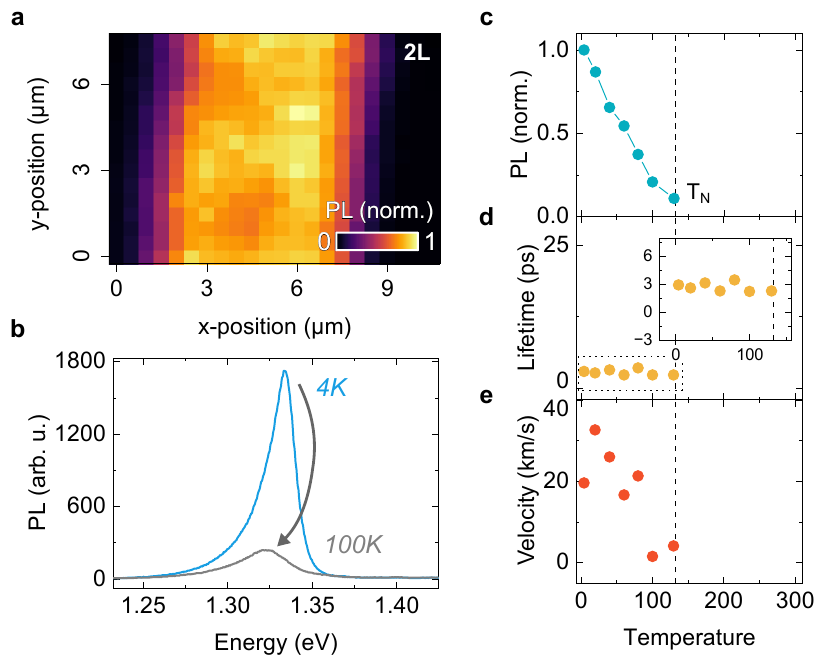}
	\caption{\textbf{Analysis of exciton properties in a 2L.} 
		\textbf{a}~Spatial PL map recorded at T = 4\,K. 
		\textbf{b}~Comparison between PL spectra measured at 4\,K and at 100\,K. 
		\textbf{c}~Integrated total PL emission as a function of temperature. 
		\textbf{d}~PL lifetime determined by a single exponential fit as a function of temperature. 
		\textbf{e}~Exciton propagation velocity as a function of temperature. 
		\label{fig:Fig-R7}}
\end{figure*}	

\begin{figure*}[h!]
	\includegraphics[scale=0.8]{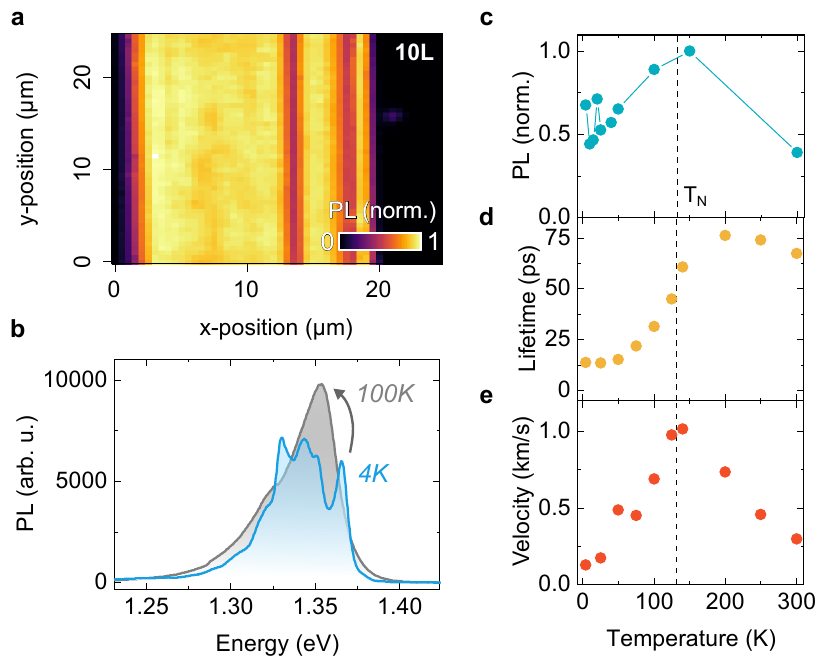}
	\caption{\textbf{Analysis of exciton properties in a 10L.}
		\textbf{a}~Spatial PL map recorded at T = 4\,K.
		\textbf{b}~Comparison between PL spectra measured at 4K and at 100K. 
		\textbf{c}~Integrated total PL emission as a function of temperature.
		\textbf{d}~PL lifetime determined by a single exponential fit as a function of temperature. 
		\textbf{e}~Exciton propagation velocity as a function of temperature. 
		\label{fig:Fig-R8}}
\end{figure*}	

\clearpage
\pagebreak

\section{Discussion of other exciton transport mechanisms}

Here we discuss the exciton and magnon transport in CrSBr from a general theoretical standpoint. Based on the experimentally observed spatial, $\sim$~$\mu$m, and temporal, $\sim 10\ldots 100$~ps, scales of exciton propagation we conclude that the most appropriate description of the propagation of quasiparticles is the drift-diffusion model. It is because the spatial and temporal scales exceed microscopic lengths and times related to the mean free path, scattering time, de Broglie wavelength, etc. of excitons. We start with the linear diffusion model in Sec.~\ref{subsec:lin}, then analyze possible nonlinear effects within the excitonic system including the exciton-exciton annihilation (Auger-like effect, cf. \cref{subsec:auger}), and Seebeck effect related to heating of the exciton gas (cf. \cref{subsec:seeb}). 
Then, we address the intertwined propagation of excitons and magnons in \cref{subsec:magn}. 

\subsection{Linear exciton diffusion}\label{subsec:lin}

The diffusion equation for the exciton density $n(\bm r, t)$, where $\bm r=(x,y)$ is the in-plane position vector and $t$ is time, for an anisotropic van der Waals semiconductor can be written as
\begin{equation}
\label{linear:diff}
\frac{\partial n}{\partial t} + \frac{n}{\tau} = \left(D_{xx} \frac{\partial^2}{\partial x^2}  + D_{yy} \frac{\partial^2}{\partial y^2} \right) n.
\end{equation}
Here we assumed that $x$ and $y$ are the principal axes in the sample plane along the crystallographic $a$-- and $b$--direction, $\tau$ is the exciton lifetime related to the radiative and non-radiative recombination processes and $D_{xx}$, $D_{yy}$ are the exciton diffusion coefficients. Microscopically, the diffusion coefficients are related to the velocity autocorrelation functions~\cite{Chernikov:2023ab,Glazov:2022aa} in the form
\begin{equation}
\label{Daa}
D_{\alpha\alpha} = \int_0^\infty \langle v_\alpha(t) v_\alpha(0)\rangle dt,
\end{equation}
where $\alpha=x,y$ is a Cartesian subscript and $v_\alpha$ is the velocity component. Based on the observed values of diffusivity we assume semiclassical exciton propagation where the autocorrelation function takes the form 
\begin{equation}
\label{vaa}
\langle v_\alpha(t) v_\alpha(0)\rangle = \langle v_\alpha^2(0)\rangle e^{-t/\tau_s},
\end{equation}
where $\tau_s$ is the scattering time and $ \langle v_\alpha^2(0)\rangle  = k_B T/M_{\alpha}$ with $k_B$ being the Boltzmann constant, $M_x$ and $M_y$ being exciton translational masses along the principal axes of the structure.

As a result, for the linear diffusion coefficients we have 
\begin{equation}
\label{Daa:1}
D_{xx}  = \frac{k_B T}{M_x} \tau_s, \quad D_{yy}  = \frac{k_B T}{M_y} \tau_s.
\end{equation}
This result also follows from the kinetic equation, see Sec.~\ref{subsec:seeb}. Equation~\eqref{Daa:1} shows that the linear exciton diffusivity is strongly anisotropic since $M_x \ne M_y$ in CrSBr.

To obtain further insight into the linear exciton diffusion scenario we estimate the diffusion coefficients~\eqref{Daa:1} based on the bandstructure from Ref.~\cite{Klein2022-1} which yields $M_x = 10.2 m_0$, $M_y = 0.6 m_0$ and $\tau_s$ extracted following Ref.~\cite{Wagner2021} from the temperature dependence of the exciton linewidth broadening $\Gamma_{exp}$ assuming the relation 
\begin{equation}
\label{Gamma}
\Gamma_{exp}=\tau_s^{-1},
\end{equation} which is valid for a number of relevant scattering mechanisms, particularly in the case of long-wavelength acoustic phonon scattering.

\begin{figure*}[t!]
	\includegraphics[scale=0.85]{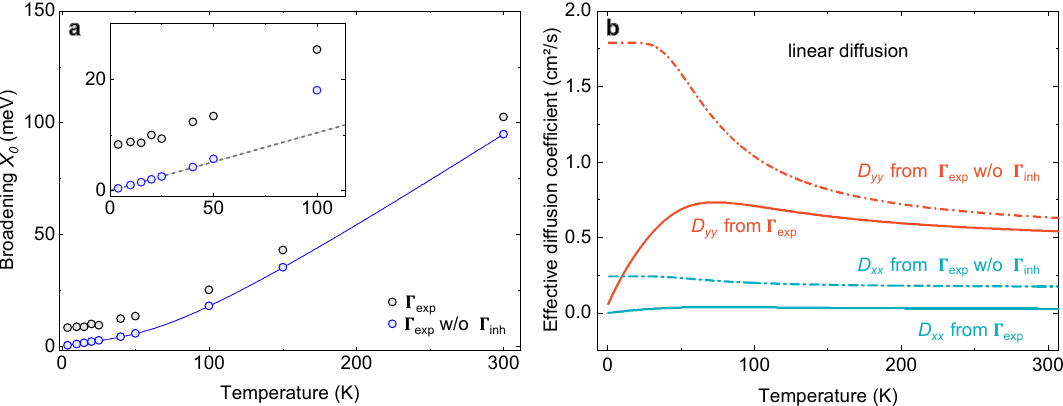}
	\caption{\textbf{Estimated temperature dependence of linear diffusion coefficients.} 
	\textbf{a}~Spectral broadening of the $X_0$ emission peak determined from temperature-dependent PL measurements (black circles). Deconvolution removes contributions from inhomogeneous broadening $\Gamma_{inh}$ (blue circles). Solid line represents the fit described in the text. Inset shows $X_0$ line broadening in the low-temperature range.
	\textbf{b}~Effective diffusion coefficients estimated from \cref{Daa:1} for different broadening parameters $\Gamma$ and crystallographic directions ($a$: turquoise, $b$: red). 		
	\label{fig:SI-Fig-1}}
\end{figure*}

The experimentally determined broadening is plotted in \cref{fig:SI-Fig-1}a. 
We fit $\Gamma_{exp}$ by a standard model, $\Gamma_{exp}(T)=\Gamma_0+c_1 T + c_2/\exp{[(E_{ph}^*/k_B T)-1]}$, with $c_{1,2}$ scaling parameters, $\Gamma_0$ a temperature independent parameter, and $E_{ph}^*$ as an effective average phonon energy. 
The contribution $\propto c_1 $ effectively accounts for the exciton interaction with long-wavelength acoustic phonons with linear dispersion and the contribution $\propto c_2$ accounts for the coupling of excitons with optical and zone edge phonons with flat dispersion.
We estimate $D_{\alpha\alpha}(T)$ for two cases of the scattering time $\tau_s(T)$.
First, we use $\tau_s^{-1}(T)=\Gamma_{exp}(T)$ to determine $D_{\alpha\alpha}(T)$ along $a$-- and $b$--directions (solid lines in \cref{fig:SI-Fig-1}b). 
This estimate assumes all processes contributing to $\Gamma_{exp}(0)$ involve momentum scattering and thus impact the diffusion of excitons. 
Second, we consider the role of inhomogeneous broadening~\cite{Wagner2021} and that momentum scattering by acoustic phonons dominates at very low temperatures.
Deconvolution of $\Gamma_{exp}(T)$ with $\Gamma_{inh}=8$\,meV renders the blue circles in \cref{fig:SI-Fig-1}a. 
The corresponding temperature dependence of the exciton diffusion coefficient is plotted as dashed dotted lines in \cref{fig:SI-Fig-1}b for transport along $a$-- and $b$--directions. 

\subsection{Exciton-exciton annihilation}\label{subsec:auger}

\begin{figure*}[h!]
	\includegraphics[scale=0.65]{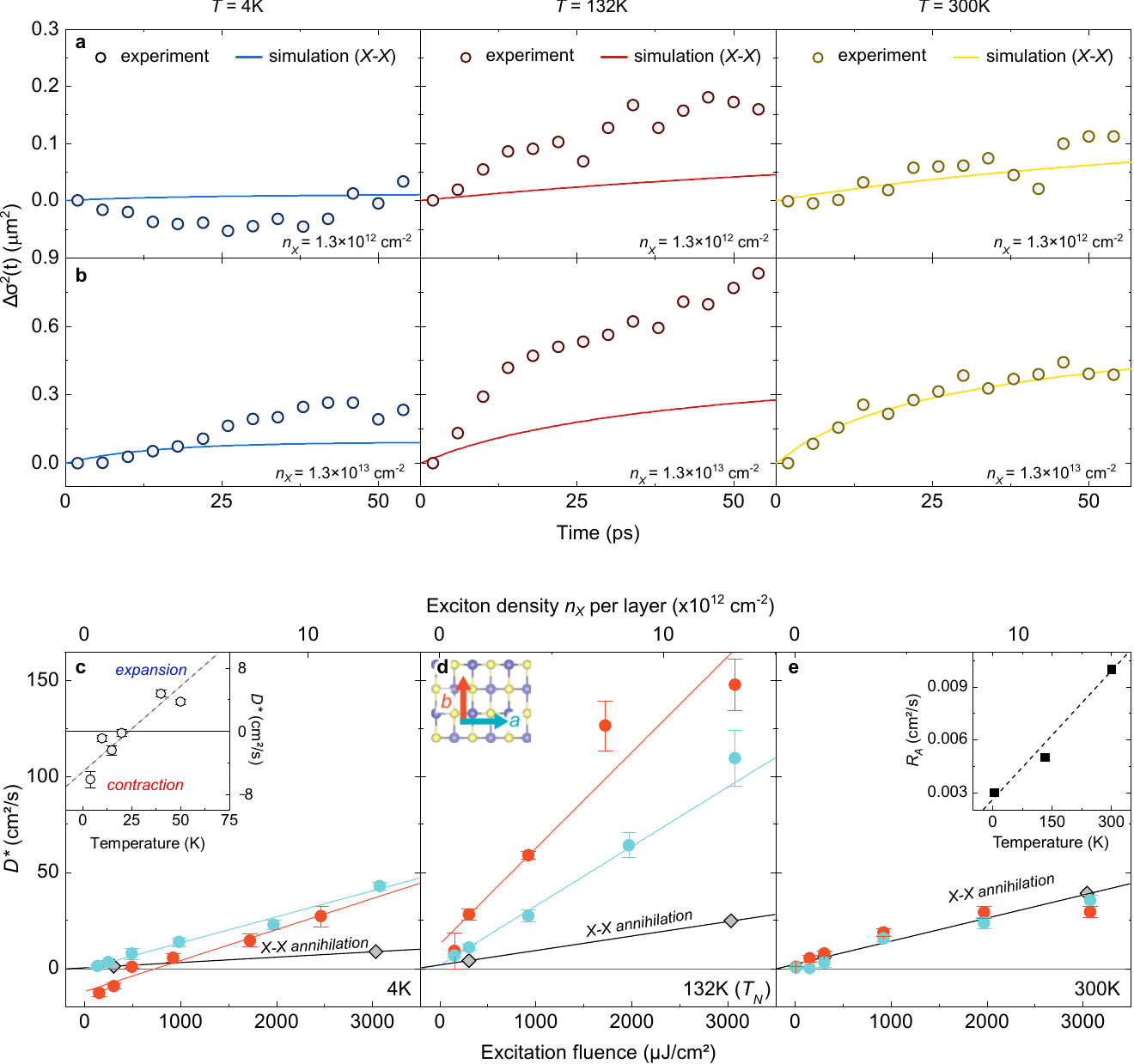}
	\caption{\textbf{Impact of X-X annihilation on exciton transport.}
		Comparison between $\Delta\sigma^2(t)$ obtained from experiments and by simulating X-X annihilation effects for an excitation fluence of \textbf{a}~310 \SI{}{\micro\J\per\square\cm} ($n_x\approx\SI{1.3e12}{\square\cm}$ per layer) and \textbf{b}~3100 \SI{}{\micro\J\per\square\cm} ($n_x\approx\SI{1.3e13}{\square\cm}$ per layer). 
		Data is shown for transport along the $b$--axis.
		Auger coefficients $R_A$ used as input for the simulation are determined directly from excitation-induced changes of the initial exciton decay rate (see inset of \textbf{e} and Section 4B for details).
		\textbf{c}-\textbf{e}~Fluence dependence of $D^*$ measured for transport along $a$-- and $b$--directions at 4\,K, 132\,K, and 300\,K. Solid lines indicate linear fits. 
		Grey diamonds and black lines represent the temperature- and density-dependent contribution of X-X annihilation effects to the measured effective diffusion coefficient $D^*$. 
		Plotted values are estimates obtained from the simulation of $\Delta\sigma^2(t)$ shown in \textbf{a,b} (cf. Section 4B). 
		Inset in \textbf{c} further shows the transition from exciton contraction to expansion under increasing sample temperatures for a fluence of 310 \SI{}{\micro\J\per\square\cm}.
		We emphasize that the particular value of the transition at about 25\,K is density dependent and thus does not represent any specific temperature.
		\label{fig:ED-Fig.9}}
\end{figure*}

Diffusion Eq.~\eqref{linear:diff} assumes linear, monomolecular exciton recombination. At elevated exciton densities, bimolecular exciton recombination processes become important. In particular, exciton-exciton (X-X) annihilation where, as a result of the collision, one exciton non-radiatively recombines and the second one takes the corresponding energy and momentum~\cite{Kumar2014,Mouri2014,Sun2014,Yuan2017}. Such processes are generally described by a non-linear, quadratic recombination term~\cite{Chernikov:2023ab},
\begin{equation}
\label{non-linear:diff}
\frac{\partial n}{\partial t} + \frac{n}{\tau} + R_A n^2 = \left(D_{xx} \frac{\partial^2}{\partial x^2}  + D_{yy} \frac{\partial^2}{\partial y^2} \right) n,
\end{equation}
and the coefficient $R_A$. 
Following Refs.~\cite{Kulig2018,Wietek2024}, we obtain an increase of effective diffusion coefficients in the form
\begin{equation}
\label{Daa:RA}
D_{{\rm eff},\alpha\alpha} = D_{\alpha\alpha} +  R_A \frac{N_0}{8\pi},
\end{equation}
where the initial exciton distribution is taken in the Gaussian form
\begin{equation}
\label{initial}
n(\bm r,0 ) = \frac{N_0}{\pi r_0^2} e^{-r^2/r_0^2},
\end{equation}
with $N_0$ being the number of injected excitons while $r_0$ determines the size of the excitation spot. We stress that the enhancement of $D_{{\rm eff},\alpha\alpha}$ is mainly related to the modification of the shape of the exciton profile which becomes more flat at the top due to an efficient increase of the recombination rate at the spot center with the highest $n(\bm r,t)$. An important consequence of this result is the fact that the enhancement of the effective diffusivity is isotropic regardless of the strong anisotropy of the linear diffusion coefficients.

\begin{figure*}[h!]
	\includegraphics[scale=1]{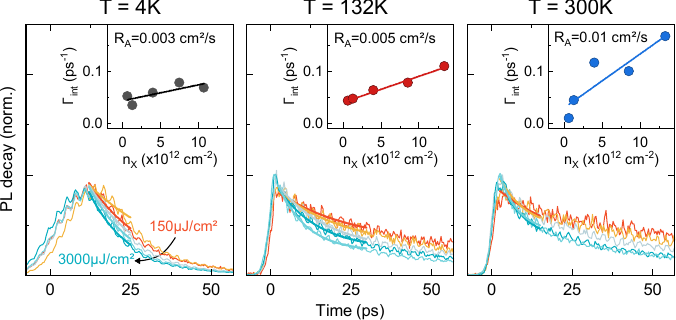}
	\caption{\textbf{Impact of X-X recombination on exciton transport.} 
		Exciton decay under increasing excitation fluence at $T=4$\,K, 132\,K, and 300\,K. Fluence-dependent changes in the initial decay rate $\Gamma_{int}$ are approximated by exponential fits. Insets: Dependence of the respective initial decay rate on exciton density $n_X$ per layer. Line represents a fit to the data, the slope of which determines the annihilation coefficient $R_A$.
		\label{fig:SI-Fig-2}}
\end{figure*}

From our time-resolved PL measurements, we can directly estimate the impact of X-X annihilation in our experiments. 
\Cref{fig:SI-Fig-2} shows the changes in the exciton decay dynamics we observe at $T=4$, 132, and 300\,K when increasing the excitation fluence.
The slope of the fluence-dependent increase of the initial decay rate $\Gamma_{int}$ directly determines the annihilation coefficients $R_A$ displayed in the inset. 
Using these coefficients as input allows us to model the temporal evolution of $\Delta\sigma^2$ under the impact of X-X annihilation by the non-linear transport equation \cref{non-linear:diff}.
In \cref{fig:ED-Fig.9}a,b, we directly compare the simulated results obtained for two different exciton densities with the corresponding values of $\Delta\sigma^2$ measured in our experiments. 
For the simulation, we use $r_0=\sqrt{2}\sigma_0\approx\SI{0.6}{\um}$ for the width of the initial exciton distribution, which we determined by imaging the laser profile on our CCD detector. 
For the exciton densities, we exemplarily use $n(0,0)=\SI{1.3E12}{cm^{-2}}$ and $n(0,0)=\SI{1.3E13}{cm^{-2}}$, which in our experiments corresponds to an excitation fluence of 310\,$\mu$J/cm$^2$ and 3100\,$\mu$J/cm$^2$, respectively.

The grey diamonds plotted in \cref{fig:ED-Fig.9} are obtained by imposing a linear fit onto the non-linear dynamics of the simulated $\Delta\sigma^2(t)$ values in the first 50\,ps, similar to the analysis of the corresponding experimental data sets. 
This facilitates a meaningful comparison between the effective diffusion coefficients obtained from the simulated and the measured values of $\Delta\sigma^2(t)$.

As can be seen in \cref{fig:ED-Fig.9}, the simulated values of $\Delta\sigma^2(t)$, which include the contribution from X-X annihilation, lead to a non-linear increase as a function of time.
At $T=4$\,K and 132\,K, we find a significant discrepancy between the simulated and measured values of $\Delta\sigma^2(t)$, indicating that X-X annihilation does not dominate the experimentally observed exciton transport dynamics. 
Moreover, the annihilation coefficient increases monotonously with temperature and does not exhibit a maximum at the Neél temperature, in particular.
We thus conclude that X-X annihilation is not responsible for the signatures of exciton transport, such as the lack of in-plane anisotropy, observed in our experiments at low and moderate $\lesssim 150$~K temperatures. 
At $T=300$\,K, however, good agreement between the measured and the modeled values of $\Delta\sigma^2(t)$ implies a major impact of the X-X annihilation effect on the transport dynamics of excitons measured in our experiments. 

Besides the influence of X-X annihilation on exciton transport, we also note that interactions between optically created electron-hole pairs can result in their repulsion and an effective increase of the observed exciton diffusion coefficient. 
This effect, however, is expected to be anisotropic due to the Einstein relation between the exciton mobility and the diffusion coefficients, thus inheriting the anisotropy of $D_{\alpha\beta}$ (further discussion provided below).

\subsubsection*{Discussion of key experimental results in the context of exciton-exciton annihilation.}

Exciton-exciton annihilation effects are well-known to dominate exciton diffusion coefficients in many systems under elevated excitation conditions, as was found to be the case in several earlier studies of exciton transport in monolayer transition metal dichalcogenides~\cite{Kulig2018,Zipfel2020,Goodman2020}.
Below, we therefore discuss point by point that key exciton transport signatures in CrSBr cannot be explained by exciton-exciton annihilation effects. 

\begin{enumerate}
\item \textbf{Maximum of $D_{eff}$ near the Néel temperature.} This is one of the central results of our study and a robust observation that is confirmed by measurements using two different excitation energies (at 1.61\,eV, off resonance, and at 1.77\,eV, in resonance with $B$-excitons). 
We find that $D_{eff}$ exhibits a maximum near $T_N$ not only at large fluence but in fact across a vast range of laser excitation fluences. 
This maximum is even observed at the smallest fluences applied in our work (see, e.g., Extended Data Figs. 4 \& 8). 
It cannot be explained by Auger recombination effects, since the extracted Auger rates substantially underestimate the resulting effective diffusion coefficients but also do not exhibit a maximum at the Néel temperature. 
Instead, they show a smooth temperature dependence, somewhat similar to that in transition metal dichalcogenides monolayers.

\item \textbf{Exciton contraction \& anisotropy at low fluence.} At low temperature and for a sufficiently small fluence, we consistently observe a contraction of the exciton cloud. 
This phenomenon was measured in 2L, 4L, 10L \& 11L thick crystals. 
Moreover, exciton transport is anisotropic under these conditions, because contraction is only prominently observed along the crystallographic $b$-direction. 
With respect to these observations, Auger recombination can neither account for contraction nor the resulting anisotropy, while the drag of excitons by magnons with negative group velocity may provide a reasonable explanation, as discussed in detail in Supplementary Section S7.

\item  \textbf{2L superdiffusion.} The superdiffusive behavior of excitons in 2L crystals is a robust phenomenon observed in 2L sampes both with and without hBN encapsulation. 
In contrast, Auger recombination would lead to sub-diffusive propagation dynamics with faster initial expansion that slows down over time as the exciton density decreases. 
It thus does not provide an explanation for this phenomenon either.

\item \textbf{Fluence dependence \& nearly isotropic transport.} These observations can result from Auger recombination. 
Indeed, both the fluence dependence of $D_{eff}$ and the observation of (nearly) isotropic transport can be signatures of Auger effects. 
However, even with respect to fluence dependence itself, Auger recombination quantitatively fails to account for the majority of experimental conditions explored in our study. 
This is especially clear for transport at the Néel temperature: Here, we observe the strongest fluence-dependence of $D_{eff}$ and almost isotropic exciton transport. 
In order to estimate whether Auger effects can be responsible for the observed effects, we analyze the changes in the initial PL decay dynamics induced by increasing fluence (cf. \cref{fig:SI-Fig-2}). 
Excitation density dependent increase of the initial rate provides an independent, quantitative evaluation of the Auger recombination (see also discussion above). 
However, the magnitude of the effect consistently fails to explain the dependence of $D_{eff}$ observed along $a$-- and $b$--directions by more than one order of magnitude. 
This strongly contrasts the results of similar analysis directly comparing density dependent recombination and diffusion in non-magnetic 2D semiconductors~\cite{Kulig2018,Zipfel2020}. 
In the case of CrSBr, to explain the experimental observation, the initial decay rate would need to be as short as one picosecond, which is clearly not within the margins of error in \cref{fig:SI-Fig-2}. 
Similar arguments can be made for our results at low temperature ($T$ = 4\,K). 
Only at room temperature, we find that Auger effects can strongly contribute or may even determine the exciton transport observed in our experiments. 
For all other conditions, this is not the case. 

\item \textbf{Linear fluence-dependence of PL intensity.} Lastly, we emphasize that our transport measurements are performed in a range of fluences characterized by a primarily linear PL response. 
This is shown in Fig.~S4 for the example of our 10L crystal. 
Only for the largest fluence values we observe a small deviation of the PL intensity from a linear response. 
This provides further evidence that Auger recombination effects are not dominating the PL decay, especially at small to intermediate fluence levels. 	
\end{enumerate}

\subsection{Exciton Seebeck effect}\label{subsec:seeb}

Excess energy stored in the course of X-X annihilation results in local heating of excitons and thus energy gradients that can drive excitons. 
These gradients are inhomogeneous in space, because they are proportional to $n(\bm r,t)$.
To analyze their impact, we consider a minimum model assuming that excitons can be described by a local temperature $T(\bm r)$ and a local chemical potential $\mu(\bm r)$, i.e., we assume that thermalization within the exciton ensemble is comparable to, or faster than, both the propagation timescales and the energy relaxation time (the latter describes the exchange of energy between excitons and the lattice). The temperature and chemical potential gradients result in the directed flux of excitons. To describe the effects microscopically we use the kinetic equation for the exciton distribution function $f\equiv f(\bm k, \bm r, t)$ in the form
\begin{equation}
\label{kinetic}
\frac{\partial f}{\partial t} + \bm v_{\bm k} \cdot \frac{\partial f}{\partial \bm r}  = Q\{f\},
\end{equation}
where the exciton velocity $\bm v_{\bm k}$ is
\begin{equation}
\label{velocity}
\bm v_{\bm k} = \frac{1}{\hbar} \frac{\partial \varepsilon_{\bm k}}{\partial \bm k} = \left(\frac{\hbar k_x}{M_x},  \frac{\hbar k_y}{M_y}\right),
\end{equation}
 and $\varepsilon_{\bm k} = \hbar^2/2(k_x^2/M_x + k_y^2/M_y)$ is the exciton dispersion. In Eq.~\eqref{kinetic}, $Q\{f\}$ is the collision integral.

We consider an exciton gas in quasi equilibrium whose chemical potential $\mu$ and temperature $T$ are smooth (on the scale of the mean free path) functions of the coordinate. Taking the exciton distribution in the form of the quasi-equilibrium one and a small non-equilibrium correction 
\begin{equation}
\label{f:quasi}
f(\bm k, \bm r,t) = \exp\left(\frac{\mu(\bm r) - \varepsilon_{\bm k}}{k_B T(\bm r)} \right) + \delta f,
\end{equation}
we obtain the following equation for the non-equilibrium correction $\delta f$
\begin{equation}
\label{df:kinetic}
-\bm v_{\bm k} \cdot \left(\bm \nabla \mu + \frac{\varepsilon_{\bm k} - \mu}{T} \bm \nabla T  \right) f_0' = -\frac{\delta f}{\tau_s},
\end{equation}
where $f_0'$ is the energy derivative of the equilibrium exciton distribution function and we used the relaxation time approximation $\tau_s$ for describing exciton scattering. 
For an anisotropic dispersion, the scattering could be anisotropic as well, but this effect is typically weaker than the effect of the mass anisotropy on transport parameters, see, e.g.,~\cite{PhysRev.101.944,gantmakher87}. 
It follows from Eq.~\eqref{df:kinetic}
\begin{equation}
\label{df:sol}
\delta f = \bm v_{\bm k} \cdot \left(\bm \nabla \mu + \frac{\varepsilon_{\bm k} - \mu}{T} \bm \nabla T  \right) f_0' \tau_s,
\end{equation}
and the exciton flux density reads (normalization volume is set to unity)
\begin{equation}
\label{i:exc}
\bm i = \sum_{\bm k} \bm v_{\bm k} \delta f = \sum_{\bm k} \bm v_{\bm k} \tau_s \left[(\bm v_{\bm k}  \cdot \nabla \mu) + (\bm v_{\bm k} \cdot \bm \nabla T) \frac{\varepsilon_{\bm k} - \mu}{T}\right] f_0'.  
\end{equation}
For non-degenerate excitons, $\mu<0, |\mu| \gg k_B T$, Eq.~\eqref{i:exc} can be further simplified to 
\begin{equation}
\label{i:exc:1}
i_\alpha = - S_{\alpha\beta} \left[-\frac{T}{\mu} \nabla_\beta \mu  +  \nabla_\beta T\right], \quad S_{\alpha\beta} = \frac{\mu}{T}\sum_{\bm k} v_{\alpha, \bm k}v_{\beta, \bm k} \tau_s f_0'.  
\end{equation}
Here $S_{\alpha\beta}$ is the tensor of Seebeck coefficients. We recall that by definition the tensors of diffusion and Seebeck coefficients are
\[
i_\alpha = - D_{\alpha\beta} \nabla_\beta n - S_{\alpha\beta} \nabla_\beta T.
\]
The density of excitons is related to their chemical potential as
\begin{equation}
\label{n:mu}
n = \frac{\sqrt{M_x M_y}}{2\pi \hbar^2} k_B T e^{\mu/k_B T} \quad \Leftrightarrow \quad \mu = k_B T\ln\left(\frac{2\pi\hbar^2 n}{k_B T\sqrt{M_x M_y}} \right),
\end{equation}
and the exciton compressibility is 
\begin{equation}
\mathcal C= \frac{\partial n}{\partial \mu} = \frac{n}{k_B T}.
\end{equation}

As a result, we obtain the diffusion coefficients in the form of Eq.~\eqref{Daa:1}
and Seebeck coefficients $S_{\alpha\beta} = (-\mu/T) \mathcal C D_{\alpha\beta}$:
\begin{equation}
\label{S}
S_{xx} = -\frac{\mu}{M_x} \frac{n}{T}\tau_s, \quad S_{yy} =- \frac{\mu}{M_y} \frac{n}{T} \tau_s.
\end{equation}
Note that the anisotropy of the diffusion and Seebeck coefficients is mainly inherited from the effective mass anisotropy.
As a result, the enhancement of exciton propagation via the Seebeck effect is expected to be strongly anisotropic: $S_{xx}/S_{yy} \propto M_y/M_x$.

\clearpage
\pagebreak

\section{Calculation of the magnon dispersion}

\begin{figure*}[h!]
	\includegraphics[scale=1.0]{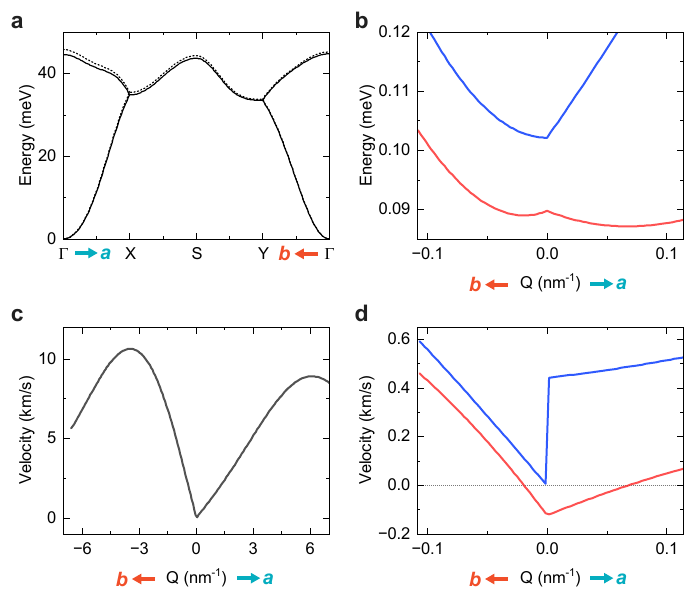}
	\caption{\textbf{Calculation of the CrSBr magnon dispersion including exchange and dipole-dipole coupling effects.} 
		\textbf{a}~Magnon dispersion in the full Brillouin zone. Solid and dashed lines in \textbf{a} represent the two branches plotted as red and blue lines in \textbf{b}.
		\textbf{b}~Effects of dipole-dipole coupling at small magnon momenta $Q$. Red and blue lines denote the two lowest branches of the dispersion. 
		\textbf{c}~Magnon velocity for large momenta in a 10L. 
		\textbf{d}~Magnon velocity for very small momenta. 
		\label{fig:SI-Fig-R10}}
\end{figure*}

\subsection{Short-range exchange interaction and basic approach}\label{subsec:short}

We consider a bilayer structure for illustration of the method, but note that for a quantitative comparison with an experimental bilayer, one needs to include the impact of surface- and substrate-related changes on the magnetic properties, or other effects. 
The Hamiltonian of magnetic system (Cr ions) in the lattice representation consists of several contributions:
\begin{itemize}
	\item The intralayer ferromagnetic interaction has the form
	\begin{equation}
	\mathcal{H}^{FM} = \sum_{i,j}J_{\langle ij\rangle}\left(\bm S^{(1)}_i\bm S^{(1)}_j + \bm S^{(2)}_i\bm S^{(2)}_j\right).
	\label{FMham}
	\end{equation}
	Accordingly, the Fourier transform of the ferromagnetic interaction reads (the sign is selected for convenience and in accordance with our previous notes)
	\begin{equation}
	J^{FM}(\bm k) \equiv -\sum_{j-i}{J}_{\langle j-i\rangle}e^{{\rm i}\bm k\bm r_{j-i}}.
	\end{equation}
	Here $i,j$ enumerate lattice sites (Cr ions), $\bm r_{j-i} \equiv \bm r_j - \bm r_i$. Note that the ferromagnetic interaction is isotropic.
	
	\item The interlayer exchange interaction (we consider it only for the sites one on top of the other)
	\begin{equation}
	\mathcal{H}^{AFM} = \sum_iJ_{int}\bm S^{(1)}_i\bm S^{(2)}_i.
	\end{equation}
	The antiferromagnetic interaction in a reciprocal space takes the form
	\begin{equation}
	J^{AFM}(\bm k) \equiv -J_{int}.
	\end{equation}
	The antiferromagnetic interaction is also isotropic in the spin space.
	
	\item The contribution describes crystal anisotropy with the easy $y$-axis
	\begin{equation}
	\mathcal{H}^{a} = \sum_iA_x\left(S^{(1)}_{i,x}S^{(1)}_{i,x} + S^{(2)}_{i,x}S^{(2)}_{i,x}\right) + \sum_iA_z\left(S^{(1)}_{i,z}S^{(1)}_{i,z} + S^{(2)}_{i,z}S^{(2)}_{i,z}\right).
	\end{equation}
\end{itemize}

In Hamiltonians above for CrSBr the dominant $J_{\langle ij\rangle} < 0$, while $J_{int} > 0$, $A_x > 0$ and $A_z > 0$. In our model in the equilibrium the spins in the first and second layers are saturated along and opposite to the $y$-axis.

It is convenient to introduce the coefficients
\begin{subequations}
	\begin{align}
	A_{\bm k} & = 2SJ_{int}, \\
	B_{\bm k} & = 2S\left[J_{int}+A_x+A_z+J^{FM}(0) - J^{FM}(\bm k)\right], \\
	C_{\bm k} & = 2S(A_z - A_x), \\
	D_{\bm k} & = 0.
	\end{align}
\end{subequations}
and express the eigenenergies of two splitted modes as
\begin{equation}
\label{dispersion:prelim}
E^\pm_{\bm k} = \frac1{2}\sqrt{(B_{\bm k} \pm D_{\bm k})^2 - (A_{\bm k} \pm C_{\bm k})^2}.
\end{equation}
These expressions can be explicitly written in the form
\begin{subequations}
	\label{dispersion:no:dd}
	\begin{align}
	E^+_{\bm k} & = 2S\sqrt{\left[A_x + \frac{J^{FM}(0)-J^{FM}(\bm k)}{2}\right]\left[J_{int}+A_z+\frac{J^{FM}(0)-J^{FM}(\bm k)}{2}\right]}, \\
	E^-_{\bm k} & = 2S\sqrt{\left[A_z + \frac{J^{FM}(0)-J^{FM}(\bm k)}{2}\right]\left[J_{int}+A_x+\frac{J^{FM}(0)-J^{FM}(\bm k)}{2}\right]}.
	\end{align}
\end{subequations}

Equations~\eqref{dispersion:prelim} and \eqref{dispersion:no:dd} are derived using the standard Holstein-Primakoff transformation introducing the Bose operators 
\begin{subequations}
	\label{HS:AF}
	\begin{align}
	&S_-^{(1)} = \sqrt{2S} a^\dag, \quad S_+^{(1)} = \sqrt{2S} a, \quad S_y^{(1)} = S - a^\dag a, \nonumber \\ 
	&S_x^{(1)} = \frac{1}{\mathrm i} \sqrt{\frac{S}{2}}(a - a^\dag), \quad S_z^{(1)} =  \sqrt{\frac{S}{2}}(a + a^\dag),\\
	\mbox{\phantom{aaa}}
	\nonumber \\
	&S_-^{(2)} = \sqrt{2S} b, \quad S_+^{(2)} = \sqrt{2S} b^\dag, \quad S_y^{(2)} = -S + b^\dag b, \nonumber \\  
	&S_x^{(2)} = \frac{1}{\mathrm i} \sqrt{\frac{S}{2}}(b^\dag - b), \quad S_z^{(2)} =  \sqrt{\frac{S}{2}}(b + b^\dag),
	\end{align}
\end{subequations}
and applying the generalized Bogoluibov transform.

\subsection{Dipole-dipole interaction}

Now let us, in addition to the short-range exchange interaction, allow for the dipole-dipole interaction of spins. We are not interested in the effect of the dipole-dipole interactions on the ground state energy and here we study only the long-range contribution to the dispersion of magnons. Thus, for our model, it is sufficient to neglect the distance between the layers and consider a 2D lattice with the effective spins [the macroscopic approach allowing to consider multilayer bulk-like crystals is presented below in Sec.~\ref{subsec:macro}]
\begin{equation}
\bm S^{eff}_i = \bm S^{(1)}_i + \bm S^{(2)}_i.
\end{equation}

By virtue of Eqs.~\eqref{HS:AF} the effective spin operator can be expressed as
\begin{equation}
\label{Seff}
S^{eff}_x = \frac1{\rm i}\sqrt{\frac{S}{2}}\left(a - b - a^\dag + b^\dag\right), \quad S^{eff}_y = 0, \quad S^{eff}_z = \sqrt{\frac{S}{2}}\left(a + b + a^\dag + b^\dag\right).
\end{equation}

The Hamiltonian of dipole-dipole interaction between the spins has a general form~\cite{ll2_eng,glazov2018electron}
\begin{equation}
\label{Hdd_discrete}
\mathcal{H}^{dd} = -\frac{\hbar^2\gamma^2}{2}\sum_{i\neq j}\frac{3\left(\bm S^{eff}_i\cdot\left(\bm r_j - \bm r_i\right)\right)\left(\bm S^{eff}_j\cdot\left(\bm r_j - \bm r_i\right)\right) - \bm S^{eff}_i\cdot \bm S^{eff}_j|\bm r_j - \bm r_i|^2}{|\bm r_j - \bm r_i|^5},
\end{equation}
where $\gamma = 2\mu_B / \hbar (s / S)$ (where $s \approx 3.56/2$~\cite{Scheie:2022aa}) is the effective gyromagnetic ratio for both layers together.
The dipole-dipole interactions are important in the range of small wavevectors only, thus, at $\bm k\to 0$, we can treat the system as continuous. In the continuous form Hamilonian~\eqref{Hdd_discrete} transforms to
\begin{equation}
\label{Hdd_continuous}
\mathcal{H}^{dd} = -\frac{\hbar^2\gamma^2}{2\mathcal{A}_0^2}\times 4 \times \int d^2\bm r\int d^2\bm \rho\frac{3\left(\bm S^{eff}(\bm r)\cdot\bm \rho\right)\left(\bm S^{eff}(\bm r + \bm \rho)\cdot\bm \rho\right) - \bm S^{eff}(\bm r)\cdot \bm S^{eff}(\bm r + \bm \rho)\rho^2}{\rho^5},
\end{equation}
where $\mathcal{A}_0$ is the area of the unit cell. The additional factor $4$ is due to the presence of 2 Cr atoms in the plane of elementary cell.

Evaluating the integrals and assuming that the short-range contributions from the dipole-dipole interaction are included in the parameters of the basic Hamiltonian introduced in Sec.~\ref{subsec:short} we find that the parameters $A_{\bm k}, \ldots D_{\bm k}$ acquire contributions due to the dipole-dipole interaction in the form
\begin{subequations}
	\begin{align}
	A^{dd}_{\bm k} & = -2S\hbar u\frac{k^2 - k_x^2}{k}, \\
	B^{dd}_{\bm k} & = -2S\hbar u\frac{k^2 - k_x^2}{k}, \\
	C^{dd}_{\bm k} & = -2S\hbar u\frac{k^2 + k_x^2}{k}, \\
	D^{dd}_{\bm k} & = -2S\hbar u\frac{k^2 + k_x^2}{k}.
	\end{align}
\end{subequations}
where we introduced the constant of the dimensionality of velocity
\begin{equation}
\label{velocity}
u = {2}\frac{\pi \hbar \gamma^2}{\mathcal{A}_0}.
\end{equation}

The energies of the magnons are
\begin{subequations}
	\label{dispersion:dd}
	\begin{align}
	E^+_{\bm k} & = 2S\sqrt{\left[A_x + \frac{J^{FM}(0)-J^{FM}(\bm k)}{2}\right]\left[J_{int}+A_z+\frac{J^{FM}(0)-J^{FM}(\bm k)}{2} - 2\hbar uk\right]}, \label{E+z}\\
	E^-_{\bm k} & = 2S\sqrt{\left[A_z + \frac{J^{FM}(0)-J^{FM}(\bm k)}{2}\right]\left[J_{int}+A_x+\frac{J^{FM}(0)-J^{FM}(\bm k)}{2} + 2\hbar u\frac{k_x^2}{k}\right]} \label{E-x}.
	\end{align}
\end{subequations}
The branch $E^+_{\bm k}$ is polarized along the $z$-axis and the branch $E^-_{\bm k}$ along the $x$-axis. Note that the magnon velocity for the branch $+$ at $k \to 0$ can be evaluated as
\begin{equation}
\label{u+}
u_+ = - 2S \sqrt{\frac{A_x}{A_z+J_{int}}} u.
\end{equation}

\subsection{Macroscopic approach to the dipole-dipole interaction}\label{subsec:macro}

\subsubsection{Magnetic susceptibility in the macroscopic approach}

We are interested at the fine structure of the magnon dispersion in the vicinity of a $\Gamma$ point. To that end we consider only homogeneous (or almost) magnetization in the layers plane. Following Refs.~\cite{ll9_eng,Akhiezer:1968aa,Gurevich2020} we present the energy density in the form
\begin{equation}
\label{macro:w}
w=J_{int}\tilde{\mathcal A}_0\left(\bm S^{(1)}\cdot \bm S^{(2)}\right) - A\tilde{\mathcal A}_0\left[\left(\bm S^{(1)}\cdot \bm n\right)^2 + \left(\bm S^{(2)}\cdot \bm n\right)^2\right] - \hbar \gamma_1 \bm h\left(\bm S^{(1)} + \bm S^{(2)} \right).
\end{equation}
Here $\tilde{\mathcal A}_0=\mathcal A_0/2$ is the magnetic unit cell area (area per one spin), $\bm S^{(i)}$ is the spin density in the layer $i$ ($i=1,2$), $\gamma_1 = -\gamma$, $\bm h$ is the magnetic field acting on the spins. Here we use the convention that the magnetization density and the spin density are related by $\bm M^{(i)} = \hbar\gamma_1 \bm S^{(i)}$. We consider, for brevity, the axially symmetric case of $A_x=A_z \equiv A>0$, taking into account that $\left(S_x^{(i)}\right)^2+\left(S_{{y}}^{(i)}\right)^2+\left(S_{{z}}^{(i)}\right)^2=const$  we have rewritten the anisotropic term in the form of Eq.~\eqref{macro:w} with $\bm n\parallel y$ being the unit vector along the $y$ (easy) axis.   The total energy $W=\int d^2r w$.

The spin dynamics equations can be derived by variational principle in the form
\begin{equation}
\label{macro:dynamics}
\frac{d\bm S^{(i)}}{d t} + \bm S^{(i)}\times \bm \Omega_i=0, \quad \bm \Omega_i = - \hbar^{-1} \frac{\delta w}{\delta \bm S^{(i)}}.
\end{equation}
Explicit calculation gives
\begin{equation}
\label{macro:fields}
\bm\Omega_i = -\frac{J_{int} \tilde{\mathcal A}_0}{\hbar} \bm S^{(\bar i)} + \frac{2A\tilde{\mathcal A}_0}{\hbar} \bm n(\bm S^{(i)}\cdot \bm n) + \gamma_1 \bm h, \quad \bar i = \begin{cases}
2~~,~~i=1,\\
1~~,~~i=2.
\end{cases}
\end{equation}
In equilibrium at $\bm h=0$ the magnetizations are aligned antiparallel: $\bm S^{(2)}_0 = - \bm S^{(1)}_0 = - S_0 \bm n$. In order to determine the magnetic susceptibility we need to find magnetization induced by $\bm h$. Let $\delta \bm S^{(i)} = \bm  S^{(i)} - \bm S^{(i)}_0$ and let $\bm m_i = \hbar \gamma_1 \delta \bm S^{(i)} \propto \bm h$. Linearizing Eqs.~\eqref{macro:dynamics} we have
\begin{subequations}
	\label{macro:dynamics:m12}
	\begin{align}
	\frac{d\bm m_1}{dt} = \frac{J_{int} \tilde{\mathcal A}_0}{\hbar}  \left[\bm m_1 \times \bm S^{(2)}_0 + \bm S^{(1)}_0 \times \bm m_2\right] - \frac{2A\tilde{\mathcal A}_0}{\hbar} \bm m_1 \times \bm S^{(1)}_0 - \hbar\gamma_1^2 \bm S^{(1)}_0 \times \bm h,\\
	\frac{d\bm m_2}{dt} = \frac{J_{int} \tilde{\mathcal A}_0}{\hbar}  \left[\bm m_2 \times \bm S^{(1)}_0 + \bm S^{(2)}_0 \times \bm m_1\right] - \frac{2A\tilde{\mathcal A}_0}{\hbar} \bm m_2 \times \bm S^{(2)}_0 - \hbar\gamma_1^2 \bm S^{(2)}_0 \times \bm h.
	\end{align}
\end{subequations}
It is convenient to introduce the total magnetization density $\bm m = \bm m_1 + \bm m_2$ and the antiferromagnetic vector density $\bm l = \bm m_1 - \bm m_2$. Hence, Eqs.~\eqref{macro:dynamics:m12} transform to 
\begin{subequations}
	\label{macro:dynamics:ml}
	\begin{align}
	\frac{d\bm m}{dt} =  - \frac{2A\tilde{\mathcal A}_0}{\hbar} \bm l \times \bm S^{(1)}_0,\\
	\frac{d\bm l}{dt} =- \frac{2J_{int} \tilde{\mathcal A}_0}{\hbar}  \bm m \times \bm S^{(1)}_0 - \frac{2A\tilde{\mathcal A}_0}{\hbar} \bm m \times \bm S^{(1)}_0 - 2\hbar\gamma_1^2 \bm S^{(1)}_0 \times \bm h.
	\end{align}
\end{subequations}
Combining Eqs.~\eqref{macro:dynamics:ml} we obtain a single equation for $\bm m$ in the form
\begin{equation}
\label{m:eq}
\frac{d^2\bm m}{dt^2} = \frac{2A\tilde{\mathcal A}_0}{\hbar}\left(\frac{2A\mathcal A_0}{\hbar}+\frac{2J_{int} \mathcal A_0}{\hbar}\right) \left[\bm m\times \bm S^{(1)}_0\right]\times \bm S^{(1)}_0 + 4A \tilde{\mathcal A}_0\gamma_1^2 \left[\bm S_0^{(1)}\times \bm h\right] \times \bm S_0^{(1)}.
\end{equation}
From Eq.~\eqref{m:eq} we obtain the magnetic susceptiblity tensor components defined by
\begin{equation}
\label{chi:2d}
m_\alpha = \chi^{2D}_{\alpha\beta} h_\beta, \quad \alpha,\beta=x,y,z,
\end{equation}
and $\bm m$ (as before) is the magnetic moment per unit area in the form
\begin{subequations}
	\label{chi:2d:comp}
	\begin{align}
	\chi^{2D}_{yy}=0, \quad \chi^{2D}_{\alpha\beta}=0~\mbox{with}~\alpha\ne \beta,\\
	\chi^{2D}_{xx} = \chi^{2D}_{zz} = \frac{F}{\Omega^2 - \omega^2},
	\end{align}
	and
	\begin{equation}
	\label{Omega}
	\Omega^2 = \frac{4}{\hbar^2}(\tilde{\mathcal A}_0 S_0)^2 A(A+J_{int}), \quad F = 4A \gamma_1^2 \tilde{\mathcal A}_0  S_0^2 = 4A \gamma_1^2 \frac{(\tilde{\mathcal A}_0 S_0)^2}{\tilde{\mathcal A}_0}.
	\end{equation}
\end{subequations}
Note that $F>0$. 

Expression~\eqref{Omega} for $\Omega$ (spin resonance frequency) is consistent with Eqs.~\eqref{dispersion:no:dd} (or Eqs.~\eqref{dispersion:dd}) at $\bm k=0$ taking into account that $S=\tilde{\mathcal A}_0 S_0 = 3/2$:
\[
E^\pm_0 = 2S \sqrt{A(A+J_{int})} = \hbar\Omega.
\]

In the vicinity of the resonance $\omega \approx \Omega$ the expressions for non-zero components of the susceptibility take the form
\begin{equation}
\label{res}
\chi^{2D}_{xx} = \chi^{2D}_{zz} \approx \frac{f}{\Omega - \omega}, \quad f = \frac{F}{2\Omega} = \gamma_1^2 \hbar \sqrt{\frac{A}{A+J_{int}}} \frac{S}{\mathcal A_0}.
\end{equation}

\subsubsection{Dipole-dipole interaction via susceptibility}

We use the macroscopic approach for accounting for the dipole-dipole interactions in magnetic system. 
To account for the dipole-dipole interactions it is sufficient to self-consistently calculate the effect of magnetic field produced by the magnetization.  We focus on the relevant case of negligble retardation, $k\gg \omega/c$ where $c$ is the speed of light, i.e., we disregard the magnons within the light cone. In this case, it is sufficient, instead of the whole set of the Maxwell equations, to solve only the magnetostatic equation
\begin{subequations}
	\label{maxwells}
	\begin{equation}
	\label{ms}
	\divv{\bm h(\bm r,t)} = -4\pi \divv{\bm m(\bm r,t)},
	\end{equation}
	with additional requirement of the curl-less field
	\begin{equation}
	\label{curlless}
	\rot{\bm h} =0.
	\end{equation}
\end{subequations}
Equation~\eqref{curlless} allows us to present $\bm h(\bm r,t) = \grad \theta(\bm r,t)$ where $\theta(\bm r, t)$ plays a role of potential.

We have for the three-dimensional magnetization density
\begin{equation}
\label{suscept:2D}
\bm m(\bm k,\omega) = \hat{\chi}^{2D}(\bm k, \omega) \bm h(\bm k, \omega)\delta(z), \quad \mbox{or} \quad m_\alpha(\bm k,\omega) = \chi^{2D}_{\alpha\beta}(\bm k, \omega)  h_\beta(\bm k, \omega)\delta(z),
\end{equation}
where the Dirac $\delta$-function accounts for the 2D nature of the studied system, $\bm k$ is the two-dimensional wavevector. Equation~\eqref{ms} for the potential $\theta$ takes the form
\begin{equation}
\label{theta:eq}
\frac{\partial^2 \theta_{\bm k}(z)}{\partial z^2} - k^2 \theta_{\bm k}(z) = 4\pi k_{\alpha}k_{\beta}\chi^{2D}_{\alpha\beta} \theta_{\bm k}(z) \delta(z) - 4\pi \frac{\partial}{\partial z} \delta(z) \chi^{2D}_{zz} \frac{\partial \theta_{\bm k}(z)}{\partial z}.
\end{equation}
Here we assumed that $\alpha,\beta$ are the in-plane Cartesian components ($x,y$), the components of the susceptibility $\chi_{\alpha,z}, \chi_{z,\alpha} \equiv 0$ and made a Fourier transform over the in-plane coordinates.

Passing to the Fourier transform along the $z$-axis, $\theta_{\bm k}(z) = \sum_{q} \theta_{\bm k,q} \exp(\mathrm i q z)$ we have from Eq.~\eqref{theta:eq}
\begin{equation}
\label{theta:eq:kq}
- (k^2+q^2) \theta_{\bm k,q} = 4\pi k_{\alpha}k_{\beta}\chi^{2D}_{\alpha\beta} \sum_{q} \theta_{\bm k,q}  + 4\pi  \chi^{2D}_{zz} q \sum_{q} q \theta_{\bm k,q}.
\end{equation}
Summing $\theta_{\bm k, q}$ from Eq. (S48) and $q\theta_{\bm k,q}$ over $q$ we obtain the equations for the eigenmodes of the system
\begin{subequations}
	\label{modes:2D}
	\begin{equation}
	\label{LT}
	1 + 2\pi \frac{k_\alpha k_\beta}{k} \chi^{2D}_{\alpha\beta} =0,
	\end{equation}
	and
	\begin{equation}
	\label{Z}
	1 - 2\pi k \chi^{2D}_{zz} =0.
	\end{equation}
\end{subequations}
Here we have made use the following expression
\[
\sum_{q} \frac{1}{k^2+q^2} = \frac{1}{2k}.
\]
In derivation of Eq.~\eqref{Z} we have also excluded the depolarization shift by the following relation (it corresponds to the inclusion of the depolarization shift into the frequency $\Omega$) 
\begin{equation}
\label{reg}
\sum_{q} \frac{q^2}{k^2+q^2} = \sum_{q} \left( 1 -  \frac{k^2}{k^2+q^2}\right) \to \sum_{q}   \frac{-k^2}{k^2+q^2} = -\frac{1}{2}k.
\end{equation}

\textbf{\emph{In-plane polarized magnon}}

Taking the susceptibility in the resonant form, Eq.~\eqref{res}, and substituting it in Eq.~\eqref{LT} we have
\begin{equation}
\label{disper:macro:LT}
\omega \equiv \Omega^*(\bm k) = \Omega + 2\pi f \frac{k_x^2}{k},
\end{equation}
where we made use of the fact that $\chi_{yy}=0$. This expression describes the dispersion of the ``longitudinal'' magnon in complete agreement with Eq.~\eqref{E-x} at $k_x \to 0$.

\textbf{\emph{Out-of-plane polarized magnon}}

In this case we have from Eq.~\eqref{Z} for the renormalized dispersion
\begin{subequations}
	\label{disper:macro:dd}
	\begin{equation}
	\Omega^*(\bm k) = \Omega -2\pi f k.
	\end{equation}
	The second term provides a negative contribution to the dispersion with the speed
	\begin{equation}
	\label{speed:macro}
	u =  - 2\pi f = - 2S \sqrt{\frac{A}{A+J_{int}}} \frac{\pi \hbar \gamma_1^2}{\tilde{\mathcal A}_0} = -{2S\sqrt{\frac{A}{A + J_{int}}}\frac{2\pi\hbar\gamma^2}{\mathcal A_0}},
	\end{equation}
\end{subequations}
in agreement with Eqs.~\eqref{E+z} and \eqref{u+}.

\subsubsection{Dipole-dipole interaction in CrSBr bulk and multilayers}

The macroscopic approach allows us to analyze the magnon dispersion for multilayer systems as well. To that end, we introduce the (bulk) magnetic susceptibility of CrSBr tensor $\hat{\chi}(\bm k, \omega)$ and consider only small wavevectors (in the vicinity of the $\Gamma$-point of the Brillouin zone) such that there are just three independent components of the susceptibility:
\begin{equation}
\label{suscept:3D}
\chi_{xx}(\bm k, \omega) = \frac{\mathcal F_{xx}}{\Omega_{x}(\bm k) - \omega}, \quad  \chi_{zz}(\bm k, \omega) = \frac{\mathcal F_{zz}}{\Omega_{z}(\bm k) - \omega}, \quad \chi_{yy}=0.
\end{equation}
The  $\chi_{yy}$ is zero because the $y$-axis is the magnetization axis. The constants $\mathcal F_{xx}, \mathcal F_{{zz}}$ determine the coupling of the spin waves with magnetic field and $\Omega_x(\bm k)$, $\Omega_{{z}}(\bm k)$ are the dispersions of the spin waves found neglecting the dipole-dipole interaction.

To find the dispersion of magnons with the dipole-dipole interactions taken into account~\cite{Keffer:1966aa,Akhiezer:1968aa,Gurevich2020} we solve the Maxwell's equations~\eqref{ms} and \eqref{curlless}. We consider a slab of  CrSBr occupying the space $-L/2 < z <L/2$ with $L$ being the slab thickness. The standard boundary conditions of continuity of tangential components of $\bm h$ and normal components of $\bm b = \bm h + 4\pi \bm m$ are imposed.

We seek the solutions of Eqs.~\eqref{maxwells} in the form 
\begin{equation}
\label{theta:z}
\theta(\bm r) = \Theta(z) e^{\mathrm i \bm k \bm r},
\end{equation}
where $\bm k = (k_x,k_y)$ is the in-plane wavevector. Our approach follows the procedure outlined in Ref.~\cite{Gurevich2020,Liu2024}.

\textbf{\emph{In-plane polarized magnons}}

To study $x$-polarized magnons let us consider the frequencies $\omega \approx \Omega_{x}$ and assume that only $\chi_{xx}$ plays a role (the magnetization is in the plane of the slab). It follows from Eq.~\eqref{maxwells} that
\begin{equation}
\label{theta:xx}
\frac{\partial^2 \Theta(z)}{\partial z^2} - k^2 \Theta(z) =
\begin{cases}
4\pi k_x^2 \chi_{xx}\Theta(z), \quad |z|<L/2,\\
0, \quad |z|>L/2,
\end{cases}
\end{equation}
where $k^2 = k_x^2 +k_y^2$.
The boundary conditions are
\begin{equation}
\label{boundary:xx}
\left.\Theta(z)\right|_{z=\pm L/2}~~\mbox{is continuous}, \quad \left.\frac{\partial\Theta(z)}{\partial z} \right|_{z=\pm L/2}~~\mbox{is continuous}.
\end{equation}
We are looking for the lowest frequency mode only and, accordingly, seek for the solution that corresponds at $\bm k\to 0$ to a homogeneous magnetization. Hence,
\begin{equation}
\label{theta:xx:sol}
\Theta(z) = \begin{cases}
\cos{(\tilde k z)}, \quad |z|< L/2,\\
\cos{(\tilde k L/2)}\exp{[-k(|z|-L/2)]}, \quad |z|>L/2,
\end{cases}
\end{equation}
where\footnote{The modes we study have frequencies in the range $\Omega_\alpha(\bm k) \leqslant \omega \leqslant \Omega_\alpha(\bm k) + 4\pi \mathcal F_{\alpha\alpha}$ where $1+4\pi \chi_{\alpha\alpha}$ is negative. As a result $\tilde k$ is real.}
\begin{equation}
\label{tilde:k:xx}
\tilde k^2 = -k^2 -4\pi \chi_{xx} k_x^2.
\end{equation}
The boundary conditions~\eqref{boundary:xx} result in the equation for the eigenmodes
\begin{equation}
\label{eigenmode:xx}
\tan{\left(\frac{\tilde k L}{2}\right)} = \frac{k}{\tilde k}.
\end{equation}
For $k\to 0$ we can find solution of Eq.~\eqref{eigenmode:xx} analytically decomposing the left- and right-hand sides in series in small $kL$ with the result
\[
\frac{\tilde k L}{2} = \frac{k}{\tilde k} \quad \Rightarrow \tilde k^2 = \frac{2k}{L}.
\]
Combining this expression with Eqs.~\eqref{suscept:3D} and \eqref{tilde:k:xx} we have
\begin{equation}
\label{Omega:x}
\omega \equiv \Omega_x^*(\bm k) = \Omega_x(\bm k) + 4\pi \mathcal F_{xx}\frac{k_x^2}{2k/L + k^2}.
\end{equation}
In particular, for $k\to 0$ we obtain
\begin{equation}
\label{Omega:x:1}
\Omega_x^*(\bm k) = \Omega_x(\bm k) + {2\pi \mathcal F_{xx}L} \frac{k_x^2}{k}.
\end{equation}
Equation~\eqref{Omega:x:1} demonstrates that the $x$-polarized magnon branch has anisotropic linear-in-$\bm k$ in the dispersion for small wavevectors in agreement with Eq.~\eqref{disper:macro:LT}. The factor $\mathcal F_{xx}L = f$ in Eqs.~\eqref{res} and \eqref{disper:macro:LT}. The absolute value of the magnon velocity at small wavevectors increases with increase in the thickness. For large $kL \gg 1$ the magnon dispersion reaches $\Omega_x^*(\bm k) \approx 4\pi \mathcal F_{xx}$, namely, the bulk longitudinal magnon frequency. True asymptotics at $kL \to \infty$ can be derived to be
\begin{equation}
\label{Omega:x:2}
\Omega_x^*(\bm k) = \Omega_x(\bm k) + 4\pi \mathcal F_{{xx}}\left[1-\left(\frac{\pi}{kL}\right)^2\right].
\end{equation}

\textbf{\emph{Out-of plane polarized magnons}}

Now we turn to the $z$-polarized magnons. To that end we consider the frequencies $\omega\approx \Omega_z$ and take into account $\chi_{zz}$ component of the susceptibility. In this way, we arrive from Eq.~\eqref{maxwells} at
\begin{equation}
\label{theta:at}
\frac{\partial^2 \Theta(z)}{\partial z^2} - k^2 \Theta(z) =
\begin{cases}
{-}4\pi \frac{\partial }{\partial z} \chi_{zz} \frac{\partial }{\partial z} \Theta(z), \quad |z|<L/2,\\
0, \quad |z|>L/2,
\end{cases}
\end{equation}
For $z$-polarized waves we have the boundary condition in the form
\begin{equation}
\label{boundary:zz}
\left.\Theta(z)\right|_{z=\pm L/2}~~\mbox{is continuous}, \quad \left.\frac{\partial}{\partial z}[1+4\pi \chi_{zz}]\Theta(z) \right|_{z=\pm L/2}~~\mbox{is continuous}.
\end{equation}
It is convenient to assume that $\chi_{zz}$ is a function of $z$ that turns to zero outside the slab.

We look for solution where $\Theta(z)$ is an odd function of the coordinate (note that $h_z = \partial \Theta/\partial z$, hence, to have an even $h_z$ the potential should be odd). Introducing
\begin{equation}
\label{tilde:k:zz}
\tilde k^2 = -\frac{k^2}{1 + 4\pi \chi_{zz}},
\end{equation}
we arrive at the following form of $\Theta(z)$:
\begin{equation}
\label{theta:zz:sol}
\Theta(z) = -\begin{cases}
\sin{(\tilde k z)}, \quad |z|< L/2,\\
\sign{z}\sin{(\tilde k L/2)}\exp{[-k(|z|-L/2)]}, \quad |z|>L/2.
\end{cases}
\end{equation}
By virtue of the boundary conditions~\eqref{boundary:zz} we obtain the equation for the eigenmodes
\begin{equation}
\label{eigenmode:zz}
\cot{\left(\frac{\tilde k L}{2}\right)} = -\frac{k}{(1+4\pi \chi_{zz})\tilde k}.
\end{equation}

Similarly to the analysis for the $x$-polarized modes, we consider the limit of $k\to 0$ (strictly speaking, $kL \ll 1$) and decomposing the cotangent for small argument arrive at
\[
\frac{2}{L}=  -\frac{k}{(1+4\pi \chi_{zz})} \quad \Rightarrow\quad 1+4\pi \chi_{zz} = - \frac{kL}{2},
\]
and using Eq.~\eqref{suscept:3D} we finally obtain
\begin{equation}
\label{Omega:z}
\omega \equiv \Omega_z^*(\bm k) = \Omega_z(\bm k) + 4\pi \mathcal F_{zz}\frac{1}{1+kL/2}.
\end{equation}
In particular, for $k\to 0$ we obtain
\begin{equation}
\label{Omega:z:1}
\Omega_z^*(\bm k) = \Omega_z(\bm k) +4\pi \mathcal F_{zz}- {2\pi \mathcal F_{zz}L} {k}.
\end{equation}
For $kL \gg 1$ we have from Eq.~\eqref{eigenmode:zz}
\begin{equation}
\label{Omega:z:2}
\Omega_z^*(\bm k) = \Omega_z(\bm k) +4\pi \mathcal F_{zz} \left(\frac{\pi}{kL}\right)^2.
\end{equation}
Equation~\eqref{Omega:z:1} shows that the $z$-polarized magnon branch has an isotropic $k$-linear contribution to the dispersion with \emph{negative} slope. Equation~\eqref{Omega:z:1} is in agreement with Eqs.~\eqref{u+} and \eqref{disper:macro:dd}.

Naturally, for $kL \ll 1$ the system is essentially two-dimensional and the dispersion is linear in the $k$. The propagation velocity linearly increases with the thickness $L$, but the range of validity of linear dispersion shrinks with increase in the thickness. The relation between the bulk and two-dimensional susceptibilities is basically the same as for the dielectric susceptibilities in two-dimensional transition metal dichalcogenides:
\begin{equation}
\label{relation:chis}
\chi^{2D}_{\alpha\beta} = \chi^{3D}_{\alpha\beta}L.
\end{equation}
Note that this expression is approximate for actual bilayer or four-layer CrSBr samples due to atomistic effects and in that case the magnetic response of a few layer CrSBr is more rigorously described within a two-dimensional model~\eqref{suscept:2D}.

\clearpage
\pagebreak

\section{Magnon-exciton drag effect}\label{subsec:magn}

We turn now towards discussing the coupled propagation of excitons and magnons. As outlined in the main manuscript, we assume that the optical excitation of the sample generates both excitons and magnons. These quasiparticle ensembles form a two-component system with mutual drag caused by the interactions of excitons and magnons. Below we discuss basic scenarios of the magnon-exciton drag effect within a semi-phenomenological model.\footnote{A detailed theory of the magnon-exciton interactions and drag based on the microscopic Hamiltonian and kinetic equation will be presented elsewhere.}

\subsection{Exciton-magnon interaction}

Microscopically, the exciton-magnon interaction can be easily understood for a two-layer case where canting of spins in the two magnetic sublattices with spins $\bm S^{(1)}$ and $\bm S^{(2)}$ results in the variation of the exciton energy, e.g., due to electron and hole interlayer tunneling in the canted configuration. In this case, intralayer excitons become mixed with the interlayer ones and their energy decreases. We present the exciton energy variation as a function of layer magnetization in the form (cf. Refs.~\cite{Wilson2021,Bae2022,Diederich2022}):
\begin{equation}
\label{V:x:m}
V = \Xi {\int d\bm r \mathcal X^\dag(\bm r)\mathcal X(\bm r)}\left[S^2 +\bm S^{(1)}(\bm r) \cdot \bm S^{(2)}(\bm r)\right],
\end{equation}
which is related to a local coupling of the intralayer exciton with the higher-in-energy interlayer one. In Eq.~\eqref{V:x:m}, $\Xi\equiv \Xi(S)<0$ is the parameter, and $\mathcal X^\dag$, $\mathcal X$ are the exciton creation and annihilation operators. The combination is 
\begin{equation}
\label{via:theta}
S^2 +\bm S^{(1)}(\bm r) \cdot \bm S^{(2)}(\bm r) = 2S^2\cos^{2}{(\theta/2)},
\end{equation}
where $\theta$ is the angle between the vectors $\bm S^{(1)}$ and $\bm S^{(2)}$. Equation~\eqref{V:x:m} demonstrates the reduction of the exciton energy for $\Xi<0$ if the magnetizations in the layers are canted with respect to their equilibrium $\bm S^{(1)} = -\bm S^{(2)}  = S \hat{\bm y}$ orientations.

It is convenient to introduce 
\begin{equation}
\label{Spm}
S_\pm^{(1,2)} = S_z^{(1,2)} \pm \mathrm i S_x^{(1,2)},
\end{equation}
obeying the commutation relations
\begin{equation}
\label{commute:pm}
S_+^{(1,2)} S_-^{(1,2)} - S_-^{(1,2)} S_+^{(1,2)} = 2S_y^{(1,2)}, \quad S_y^{(1,2)} S_\pm^{(1,2)} - S_\pm^{(1,2)} S_y^{(1,2)} = \pm S_\pm^{(1,2)}.
\end{equation}
Let $S_y^{(1)} = -S_y^{(2)} \equiv S>0$ (antiferromagnetic configuration at saturation). In the case of classically large spin we can use the Holstein-Primakoff transformation~\cite{ll9_eng} and present the spin operators in the approximate form 
\begin{subequations}
	\label{HS:AF2}
	\begin{align}
	S_-^{(1)} = \sqrt{2S} a^\dag, \quad S_+^{(1)} = \sqrt{2S} a, \quad S_y^{(1)} = S - a^\dag a, \quad S_x^{(1)} = \frac{1}{\mathrm i} \sqrt{\frac{S}{2}}(a - a^\dag), \quad S_z^{(1)} =  \sqrt{\frac{S}{2}}(a + a^\dag),\\
	S_-^{(2)} = \sqrt{2S} b, \quad S_+^{(2)} = \sqrt{2S} b^\dag, \quad S_y^{(2)} = -S + b^\dag b, \quad S_x^{(2)} = \frac{1}{\mathrm i} \sqrt{\frac{S}{2}}(b^\dag - b), \quad S_z^{(2)} =  \sqrt{\frac{S}{2}}(b + b^\dag),
	\end{align}
\end{subequations}
via bosonic operators $a$, $a^\dag$, $b$, $b^\dag$. Making use of Eq.~\eqref{HS:AF2}, we obtain (keeping only the lowest powers of the $a$ and $b$ operators)
\begin{equation}
\label{V:x:m:1}
V = \Xi S{\int d\bm r \mathcal X^\dag(\bm r) \mathcal X(\bm r)}\left[a^\dag(\bm r) a(\bm r)+ b^\dag(\bm r) b(\bm r) + a^\dag(\bm r) b^\dag(\bm r) + a(\bm r) b(\bm r)\right].
\end{equation}
Operators $a$, $a^\dag$, $b$, $b^\dag$ are linearly related (via the Bogolyubov transform) with the magnon creation and annihilation operators. Thus, Eq.~\eqref{V:x:m} shows that only two-magnon processes are of importance: exciton can be scattered by the magnon or can emit or absorb two magnons simultaneously.

To obtain deeper insight into this two-magnon-exciton interaction, let us analyze the exciton-magnon interaction from the symmetry standpoint. We focus on the coupling between a non-degenerate optically active exciton state (allowed in $y$ polarization) that transforms according to the $B_{2u}$ or $\Gamma_2^-$ irreducible representation of the $D_{2h}$ point group. The effective Hamiltonian of exciton-magnon interaction should be invariant under all point-group transformations. Since $B_{2u} \times B_{2u} = A_g$ (or $\Gamma_1^+$, identity representation) the only components or combinations of components of magnetization that enter the coupling Hamiltonian should be invariant. In our case it is $S_y$, $S_y^2$, $S_x^2$, $S_z^2$, \ldots (in each layer the magnetization is either parallel or antiparallel to the easy-$y$-axis, thus $S_y$ is invariant). These combinations are quadratic in the magnon operators, thus only two-magnon processes are possible at zero external magnetic field in agreement with the microscopic analysis presented above.

\subsection{Two component exciton-magnon drift-diffusion}\label{subsec:dr:diff}

Under the same assumptions as in \cref{subsec:lin}, let us derive phenomenological equations describing the propagation of excitons and magnons. Introducing $m(\bm r, t)$, the magnon density\footnote{Here, for simplicity, we sum over all magnon branches.}, we obtain, assuming its smooth variation in space and time, the set of continuity equations for magnons and excitons
\begin{subequations}
	\label{cont}
	\begin{align}
	\frac{\partial m}{\partial t} + \bm \nabla \cdot \bm j + R_m\{m\} = G_m,\label{cont:magnon}\\
	\frac{\partial n}{\partial t} + \bm \nabla \cdot \bm i + R_x\{n\} = G_x.\label{cont:exc}
	\end{align}
\end{subequations}
Here $\bm j$ and $\bm i$ are the magnon and exciton fluxes, $G_m$ and $G_x$ the magnon and exciton generation rates, respectively, $R_m\{m\}$ and $R_x\{n\}$ are the operators describing relaxation of the quasiparticles, e.g., $R_x\{n\} = n/\tau + R_A n^2$ describing monomolecular and bimolecular recombination of excitons.

Within the drift-diffusion approach, the fluxes should be related to the density gradients and interparticle drag forces related to exciton-magnon interaction. We focus here on the exciton flux and present it in the form
\begin{equation}
\label{exc:flux}
\bm i = \bm i_{\rm diff} + \bm i_{\rm dr},
\end{equation}
via the diffusive $i_{{\rm diff},\alpha} = - D_{\alpha\beta} \partial n/\partial x_\beta$ and $\bm i_{\rm dr}$ components. We recast the drift contribution as
\begin{equation}
\label{exc:drift}
\bm i_{\rm dr} = n \bm v,
\end{equation}
where $\bm v$ is the exciton drift velocity determined as an average over the exciton ensemble
\begin{equation}
\label{v:def}
\bm v = \sum_{\bm k} \bm v_{\bm k} f_{\bm k},
\end{equation} 
in terms of a non-equilibrium distribution function of excitons $f_{\bm k}$. Similarly, the drift flux of magnons reads
\begin{equation}
\label{magn:drift}
\bm j_{\rm dr} = m \bm u,
\end{equation}
where $\bm u$ is the magnon drift velocity.

Due to the exciton-magnon interaction the exciton and magnon velocities are coupled as
\begin{equation}
\label{phenom:v}
\frac{d v_\alpha}{dt} + \Gamma_{\alpha\beta}^0 v_\beta + \Gamma_{\alpha\beta}^m (v_\beta - u_\beta) = M_{\alpha\beta}^{-1} F_\beta.
\end{equation}
Here $\Gamma^0_{\alpha\beta}$ is the exciton momentum relaxation rates tensor unrelated to the exciton-magnon scattering (due to phonons, disorder, etc.) and $\Gamma_{\alpha\beta}^m$ is the tensor of exciton-magnon scattering rates. The rates $\Gamma_{\alpha\beta}^m$ can depend on the excitation power due to magnon population. In Eq.~\eqref{phenom:v}, $\bm F$ is the external force acting on the excitons (due to the Seebeck effect of excitons, exciton-exciton repulsion\footnote{In the presence of effective exciton-exciton repulsion $\bm F \propto \bm \nabla n$ and effective diffusion coefficient is enhanced. However, this enhancement is anisotropic due to inverse effective mass factor in the right hand side of Eq. (S30), see also Refs.~[1,7] for details.} or phonon drag/wind), see Refs.~\cite{keldysh_wind,kozub_drag,Bulatov1992,1996JETP...82.1180K,PSSB:PSSB45,Glazov2019,Perea-Causin2019} for details.

Assuming that $\bm F=0$ and that $x$ and $y$ are the principal axes of the structure, we have in the steady state
\begin{equation}
\label{v:steady}
v_\alpha = u_\alpha \frac{\Gamma_{\alpha\alpha}^m}{\Gamma_{\alpha\alpha}^0+\Gamma_{\alpha\alpha}^m}, \quad \alpha=x~~\mbox{or}~~y.
\end{equation}
At $\Gamma^m \ll \Gamma^0$ (conventional, linear drag regimes), the exciton velocity $v\ll u$. In this case only a small fraction of magnon momentum is transferred to excitons. At $\Gamma^m \gg \Gamma^0$ (a lot of magnons, highly nonlinear regime) the exciton velocity saturates at $\bm u$ irrespectively of its direction. Thus, the propagation speed of excitons is limited by the magnon speed.

Equation~\eqref{v:steady} is the central result of our model. It allows to understand key features of the exciton propagation in few-layer CrSBr:
\begin{enumerate}
	
	\item Fast exciton propagation with effective diffusion coefficients exceeding by far the estimates based on the exciton linear diffusion model in Sec.~\ref{subsec:lin}. Indeed, we assume that for a wide range of excitation powers used in experiment the magnon occupancies are high and exciton-magnon scattering dominates over the exciton momentum relaxation related to phonon and disorder scattering:
	\begin{equation}
	\label{eff:cond}
	\Gamma^m \gtrsim \Gamma^0.
	\end{equation}
	In that case exciton propagation velocities are close to those of magnons. 
	The exciton transport velocities ($\sim$km/s) observed in our experiment are in overall good agreement with the transport velocities of magnons reported in recent studies~\cite{Bae2022}. 
	
	\item Maximum of the effective exciton diffusion coefficient in the vicinity of the Néel temperature $T_N$: An increase in the magnon occupancies as the temperature $T$ approaches $T_N$ results in the increase of $\Gamma^m$. As a result, the effective diffusion coefficient of excitons should have a maximum at $T\approx T_N$.
	
	\item Almost isotropic exciton propagation:  Under condition~\eqref{eff:cond}, the excitons co-propagate with magnons whose dispersion is just weakly anisotropic in contrast to that of excitons (cf. Refs. \cite{Bae2022} and \cite{Klein2022-1}). Hence, the magnon cloud expands virtually isotropically and due to exciton-magnon interactions, the expansion of the exciton cloud follows the magnon cloud expansion.
	
\end{enumerate}

\subsection{Contraction of the exciton cloud}

We briefly discuss two mechanisms potentially responsible for the contraction of excitons observed in our measurements at low temperature and under low excitation fluence (see, for example, Fig. 3 of the main manuscript). 
In the first mechanism, we consider that optical excitation may locally create a transient, attractive potential for excitons which leads to exciton funneling, an effect that would be observed as negative transport (contraction) in our experiments. 
From a comparison with our numerical model, however, we conclude that the intrinsic diffusion coefficients required in this model to match our experimental results are unreasonable. 
We therefore also consider that the negative group velocity of low-frequency, $\Gamma$-point magnons in CrSBr reported in a recent study~\cite{Sun2024} could lead to a negative magnon-exciton drag effect. 

\subsubsection{Excitation-induced transient attractive potential} 

With the drift-diffusion model, we estimate the transport of excitons in the presence of an attractive, excitation-induced transient potential. 
Such a potential may in principle be created by the distortion of the magnetic order in the region of optical excitation, as is discussed in the main text in the context of Fig.~2. 
It is important to note, however, that negative transport of excitons is only observed under low excitation fluence. 
The density of thermal fluctuations (incoherent magnons) induced in this case are therefore small. 
Nonetheless, they may lead to a local minimum in the exciton energy landscape. 
From optical spectroscopy, however, we can conclude that the depth of this potential cannot be more than $\approx-1$\,meV; otherwise, it would lead to notable spectral shifts of the exciton emission peak in the upper panel of Fig. 2c of the main manuscript. 

\begin{figure*}[t!]
	\includegraphics[scale=1.5]{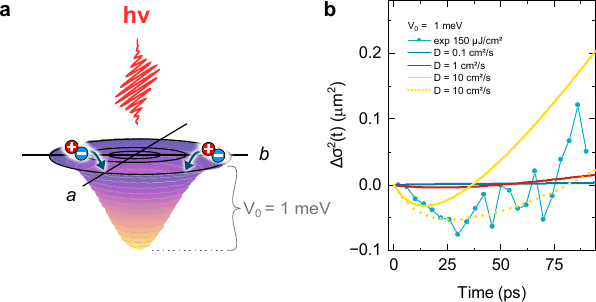}
	\caption{\textbf{Simulation of exciton transport dynamics in the presence of an attractive potential.} 
		\textbf{a}~Sketch of the attractive potential considered in the model. Spectral analysis of the $X_0$ emission peak limits the depth of the potential to $V_0\approx1$\,meV.
		\textbf{b}~Calculated time evolution of $\Delta\sigma^2$ for different values of the exciton diffusion coefficient $D$. Depth of the potential used in the simulations is $V_0$=1\,meV, the exciton lifetime is $\tau$=15\,ps, and $T$=5\,K. The decay constant of the potential is $\tau_\varepsilon=2\tau$ for all curves except for the yellow dashed line ($\tau_\varepsilon=6\tau$). Blue dots and line represent $\Delta\sigma^2$ values measured in a 10L crystal under 155 \SI{}{\micro\J\per\square\cm}.
		\label{fig:SI-Fig-4}}
\end{figure*}

We therefore model exciton transport by including in the left-hand side of kinetic Eq.~\eqref{kinetic} the force term
\begin{equation}
\label{force:kin}
\frac{\bm F}{\hbar} \cdot \frac{\partial f}{\partial \bm k}, \quad \mbox{where} \quad \bm F = - \bm \nabla V(\bm r, t),
\end{equation} 
where $V(\bm r, t)$ is the attractive potential produced by the magnon cloud. As a result, the diffusion Eq.~\eqref{linear:diff} takes the form
\begin{equation}
\label{linear:diff:potential}
\frac{\partial n}{\partial t} + \frac{n}{\tau_r} = \sum_{\alpha,\beta} D_{\alpha\beta} \left[\frac{\partial^2}{\partial x_\alpha x_\beta} n +  \frac{1}{k_B T} \frac{\partial }{\partial x_\alpha} n \frac{\partial V}{\partial x_\beta}\right] .
\end{equation}
Figure~\ref{fig:SI-Fig-4}b shows the results of a simulation of exciton transport within this model with 
\begin{equation}
\label{V:r:t}
V(\bm r, t) = - V_0 \exp\left(-\frac{r^2}{r_0^2} - \frac{t}{\tau_\varepsilon} \right),
\end{equation} 
where $r_0$ is the excitation spot size [cf. Eq.~\eqref{initial}], $V_0>0$ is the depth of the potential well created by the magnons, and $\tau_\varepsilon$ is the decay time of this potential. It is seen from Fig.~\ref{fig:SI-Fig-4}b that to describe the dynamics of $\Delta \sigma^2(t)$ for a reasonable depth of the potential well and realistic values of temperature, the diffusion coefficient should significantly exceed the value estimated on the basis of the linewidths.

\subsubsection{Negative magnon group velocity and exciton-magnon drag} 

Another possible scenario for the contraction of the exciton cloud is related to the specifics of the magnon dispersion in CrSBr in the region of small wavevectors $k \lesssim 1$~$\mu$m$^{-1}$, which are the result of dipole-dipole interaction in the spin system. Indeed, as demonstrated theoretically and experimentally~\cite{Sun2024} the group velocity of one of the magnon branches becomes negative or, strictly speaking, the group and phase velocities become antiparallel (see Fig.~\ref{fig:SI-Fig-R10} and also Fig.~3b of Ref.~\cite{Sun2024}):
\begin{equation}
\label{neg:vg}
\bm v_{g,-} = \frac{\partial\omega_-(\bm k)}{\partial \bm k} \uparrow\downarrow \bm k.
\end{equation}
As a result, two effects become possible
\begin{enumerate}
	\item The non-equilibrium magnon can contract, since magnons with negative group velocity can propagate towards the density gradient in contrast to the equilibrium diffusion where the particles propagate against the density gradient.
	\item The direction of exciton drag can be opposite to the direction of the magnon propagation as a result of the negative magnon dispersion in the energy and momentum conservation laws underlying exciton magnon collisions.
\end{enumerate}

Below we present a model which demonstrates that the exciton drag can be opposite to the direction of magnon propagation (scenario 2 above).

\subsubsection{Kinetic approach to exciton contraction}

We present the kinetic equation~\eqref{kinetic} for the exciton distribution function in the form
\begin{equation}
\label{kin:eq}
\frac{\partial f}{\partial t} + \bm v_{\bm k} \frac{\partial f}{\partial \bm r} = Q^0\{f\} + Q^m\{f,m\}, 
\end{equation}
where we explicitly separated the collision integral in two contributions $Q^0\{f\}$ and $Q^m\{f, m\}$ that describe exciton scattering by static disorder/phonons and by magnons, respectively.  Hereafter we omit all irrelevant arguments of the distribution function for brevity.

In the case of exciton-magnon scattering (2 magnon processes) the collision integral describing the process
\[
\mbox{exciton}~\bm k +~\mbox{magnon}~\bm p \to \mbox{exciton}~\bm k+\bm q +~\mbox{magnon}~\bm p-\bm q,
\]
where $\bm q$ is the transferred wavevector, reads
\begin{multline}
\label{Qm:scatt}
Q^m\{f,m\} = \sum_{\bm q,\bm p} \frac{2\pi}{\hbar}|M(\bm k, \bm p; \bm q)|^2 \delta(\varepsilon_{\bm k} + E_{\bm p} - \varepsilon_{\bm k+ \bm q} - E_{\bm p - \bm q}) \\
\times \{f(\bm k+ \bm q)[1+f(\bm k)] m_{\bm p -\bm q}[1+m_{\bm p}] -f(\bm k)[1+f(\bm k+\bm q)] m_{\bm p}[1+m_{\bm p-\bm q}]\}.
\end{multline}
Here $M(\bm k, \bm p; \bm q)$ is the appropriate matrix element, $\varepsilon_{\bm k}$ is the exciton dispersion, $E_{\bm p}$ is the magnon dispersion [for simplicity we consider only one branch],  $m_{\bm p}$ is the (nonequilibrium) magnon distribution function. Equation~\eqref{Qm:scatt} accouts for the processes where exciton with the wavevector $\bm k$ and magnon with the wavevector $\bm p$ scatter each other and, as a result, exciton arrives at the state with the wavevector $\bm k+\bm q$ and magnon to the state $\bm p- \bm q$ (and the reverse process), where $\bm q$ is the scattering wavevector.

We neglect exciton exciton collisions and assume that the magnon distribution function $m_{\bm p}$ ($\bm p$ is the magnon wavevector) is in the form
\begin{equation}
\label{m:diff}
m_{\bm p} = m_0(E_{\bm p}) + m_0'(E_{\bm p}) \tau_m (\bm V_{\bm p} \bm \nabla \mu_m).
\end{equation}
Here, the first term, $m_0(E_p)$, is the isotropic quasi-equilibrium magnon distribution function, and the second term represents the non-equilibrium contribution related to the magnon propagation, $\tau_m$ is the magnon relaxation time, $\mu_m$ is the magnon chemical potential, and $\bm V_{\bm p} = \hbar^{-1} \partial E_{\bm p}/\partial \bm p$ is the magnon velocity. We use the notation $m_0'(E_{\bm p})=dm_0/dE_{\bm p}$.
Note that the total magnon velocity described by the distribution function~\eqref{m:diff}
\begin{equation}
\label{u:diff}
\bm u = \frac{\sum_{\bm p} \bm V_{\bm p}  m_0'(E_{\bm p}) \tau_m (\bm V_{\bm p} \bm \nabla \mu_m)}{\sum_{\bm p} m_0(E_{\bm p})} = \tau_m \frac{\bm \nabla \mu}{2} \frac{\sum_{\bm p} V^2_{\bm p}  m_0'(E_{\bm p})}{\sum_{\bm p} m_0(E_{\bm p})} \uparrow\downarrow \bm \nabla \mu.
\end{equation}
The magnon flux is counter to the gradient of chemical potential, as expected ($m_0'<0$ for equilibrium distribution) regardless the sign of the magnon dispersion.

Linearizing the collision integrals and solving the kinetic equation we arrive at
\begin{multline}
\label{delta:f:1:diff}
\delta f(\bm k) 
= \tau_p\times \frac{2\pi}{\hbar} f_0'(\varepsilon_{\bm k }) \sum_{\bm p, \bm q} |M(\bm k, \bm p; \bm q)|^2\delta(\varepsilon_{\bm k} + E_{\bm p} - \varepsilon_{\bm k+ \bm q} - E_{\bm p - \bm q})m_0(E_{\bm p})\\
\times 
\tau_m\left[\left( \bm V_{\bm p-\bm q} - \bm V_{\bm p}  \right)\cdot \bm \nabla \mu_m \right],
\end{multline}
where as above we assumed that the occupancies of the states are small, $f_0'(\varepsilon_{\bm k}) = df_0/d\varepsilon$ and $\tau_p$ is the exciton momentum relaxation time contributed both by exciton-magnon and exciton-phonon/disorder collisions [cf. Eq.~\eqref{phenom:v}]
\[
\frac{1}{\tau_p} = \Gamma^0 + \Gamma^m.
\]
Hereafter we assume either isotropic approximation or one-dimensional propagation, that is why the subscripts $\alpha$ and $\beta$ are omitted.

Neglecting the magnon dispersion compared to the exciton dispersion in the energy conservation $\delta$-function, changing the summation variable from $\bm q$ to $\bm k' = \bm k + \bm q$ and assuming isotropy in the plane or one-dimensional propagation we have
\begin{equation}
\label{delta:f:2:diff}
\delta f(\bm k) 
= \tau_p\times \frac{2\pi}{\hbar} f_0'(\varepsilon_{\bm k }) \sum_{\bm p, \bm k'} |M( \bm k, \bm p; \bm k' - \bm k)|^2\delta(\varepsilon_{\bm k}  - \varepsilon_{\bm k'})m_0(E_{\bm p})\left( \bm V_{\bm p+\bm k - \bm k'}\cdot {\bm \nabla \mu_m \tau_m}\right).
\end{equation} 
The excitons drift with the velocity
\begin{multline}
\label{v:drift:diff}
\bm v = {\frac{1}{\sum_{\bm k} f_0(\varepsilon_{\bm k})}} \sum_{\bm k} \frac{1}{\hbar} \frac{\partial \varepsilon_{\bm k}}{\partial \bm k} \delta f(\bm k) \\
=   \tau_p \frac{2\pi}{\hbar}\sum_{\bm p, \bm k, \bm k'} f_0'(\varepsilon_{\bm k}) |M(\bm k, \bm p; \bm k' - \bm k)|^2\delta(\varepsilon_{\bm k}  - \varepsilon_{\bm k'})m_0(E_{\bm p})\bm v_{\bm k}\left( \bm V_{\bm p+\bm k - \bm k'}\cdot {\bm \nabla \mu_m \tau_m}\right).
\end{multline}

Equation~\eqref{v:drift:diff} can be simplified assuming that the magnon momentum $\bm p$ exceeds by far the momentum of exciton. Such a large-$p$/small-$k$ asymptotics corresponds to the regime studied above in Sec.~\ref{subsec:dr:diff}.  Hence, in Eq.~\eqref{v:drift:diff} we decompose $\bm V_{\bm p + \bm k - \bm k'}$ in series over $|\bm k - \bm k'| \ll p$. Technically, it is convenient to introduce a novel variable $\bm P = \bm p + \bm k - \bm k'$, rewrite the sum as 
\begin{multline}
\label{v:drift:diff:1}
\bm v =  {\frac{1}{\sum_{\bm k} f_0(\varepsilon_{\bm k})}}\tau_p \frac{2\pi}{\hbar}\sum_{\bm P, \bm k, \bm k'} f_0'(\varepsilon_{\bm k}) |M(\bm k, \bm p; \bm k' - \bm k)|^2\delta(\varepsilon_{\bm k}  - \varepsilon_{\bm k'})m_0(E_{\bm P + \bm k' - \bm k})\bm v_{\bm k}\left( \bm V_{\bm P}\cdot {\bm \nabla \mu_m \tau_m}\right) \\
={\frac{1}{\sum_{\bm k} f_0(\varepsilon_{\bm k})}} \tau_p \frac{2\pi}{\hbar}\sum_{\bm P, \bm k, \bm k'} f_0'(\varepsilon_{\bm k}) |M(\bm k, \bm p; \bm k' - \bm k)|^2\delta(\varepsilon_{\bm k}  - \varepsilon_{\bm k'})m_0'(E_{\bm P})\hbar [(\bm k'- \bm k) \bm V_{\bm P}]\bm v_{\bm k}\left( \bm V_{\bm P}\cdot {\bm \nabla \mu_m \tau_m}\right).
\end{multline}
This expression can be brought to the form similar to Eq.~\eqref{v:steady}: 
\begin{equation}
\label{v:drift:diff:2}
\bm v = \tau_p \Gamma^m \bm u = \frac{\Gamma^m}{\Gamma^0+\Gamma^m} \bm u,
\end{equation}
where the magnon velocity $\bm u$ is defined in Eq.~\eqref{u:diff}. Note that in this approximation
\begin{multline}
\label{delta:f:3:diff}
\delta f(\bm k) = \tau_p\times \frac{2\pi}{\hbar} f_0'(\varepsilon_{\bm k }) \sum_{\bm p, \bm k'} |M(\bm k, \bm p; \bm k' - \bm k)|^2\delta(\varepsilon_{\bm k}  - \varepsilon_{\bm k'})m_0'(E_{\bm p}) \hbar [(\bm k'- \bm k) \bm V_{\bm p}] \left( \bm V_{\bm p}\cdot {\bm \nabla \mu_m \tau_m}\right) 
\\= \tau_p\times  (-\bm k) f_0'(\varepsilon_{\bm k }) \frac{2\pi}{\hbar} \underbrace{\sum_{\bm k'} |M(\bm k, \bm p; \bm k' - \bm k)|^2\delta(\varepsilon_{\bm k}  - \varepsilon_{\bm k'}) (1-\cos{\vartheta_{\bm k,\bm k'}})}_{\Gamma^m/\sum_{\bm p} m_0(E_{\bm p})}\\
\times \underbrace{\sum_{\bm p} \tau_mm_0'(E_{\bm p}) \bm V_{\bm p}[\bm V_{\bm p} \bm \nabla \mu_m]}_{\bm u \sum_{\bm p} m_0(E_{\bm p})}.
\end{multline}

Let us now analyze the non-equilibrium exciton distribution function and drift velocity depending on the properties of exciton-magnon interaction. To that end, let us consider a simple one-dimensional version of the model above which simplifies the calculations and allows us to present the results for the distribution function in a simple analytical form. We note that the energy conservation law $\varepsilon_k = \varepsilon_{k'}$ provides two solutions: (i) $k'=k$ and (ii) $k' = - k$. The first solution does not provide any contribution to $\delta f(k)$ in Eq.~\eqref{delta:f:2:diff} after summation over $p$ since $V_p$ is an odd function of p. The remaining contribution simplifies to 
\begin{equation}
\label{delta:f:2:diff:1D}
\delta f(\bm k) = \tau_p\times \frac{2\pi}{\hbar} f_0'(\varepsilon_{ k }) \sum_{ p,  k'} |M(k,p;k' -  k)|^2\delta(\varepsilon_{ k}  - \varepsilon_{ k'})m_0(E_{ p})\left(  V_{ p+ k -  k'}\cdot {\nabla \mu_m \tau_m}\right).
\end{equation}

\begin{figure}[h]
	\includegraphics[width=0.45\textwidth]{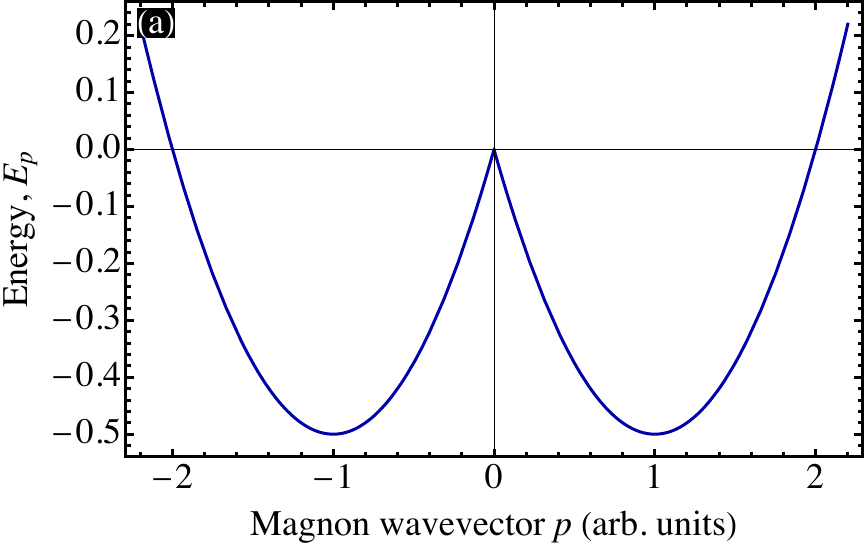}
	\includegraphics[width=0.45\textwidth]{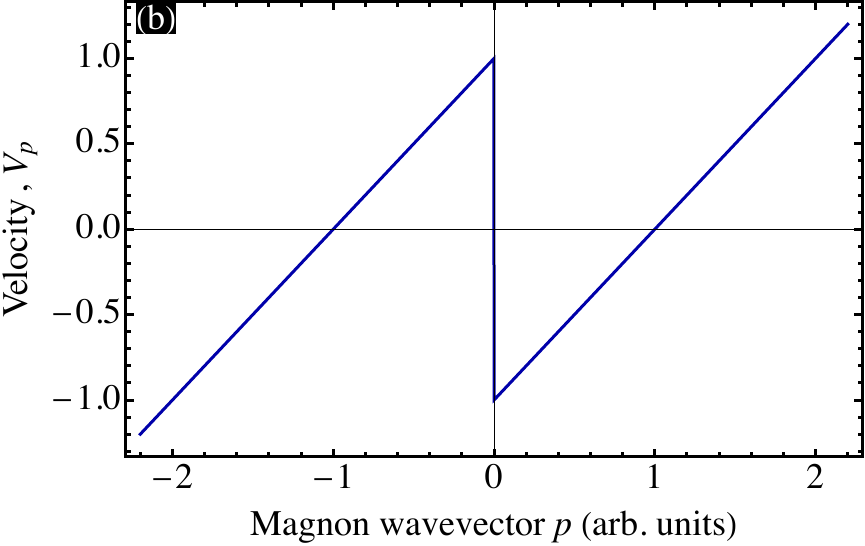}\\
	\vspace{0.5cm}
	\includegraphics[width=0.45\textwidth]{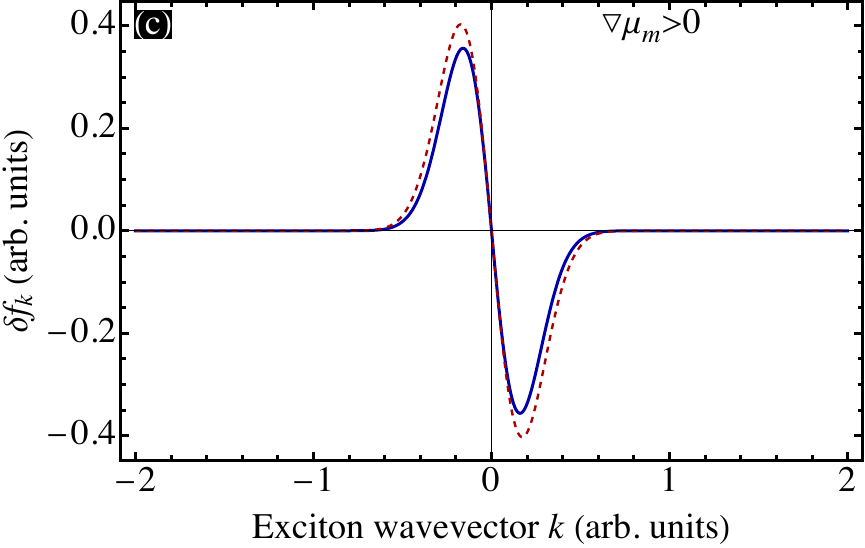}
	\includegraphics[width=0.45\textwidth]{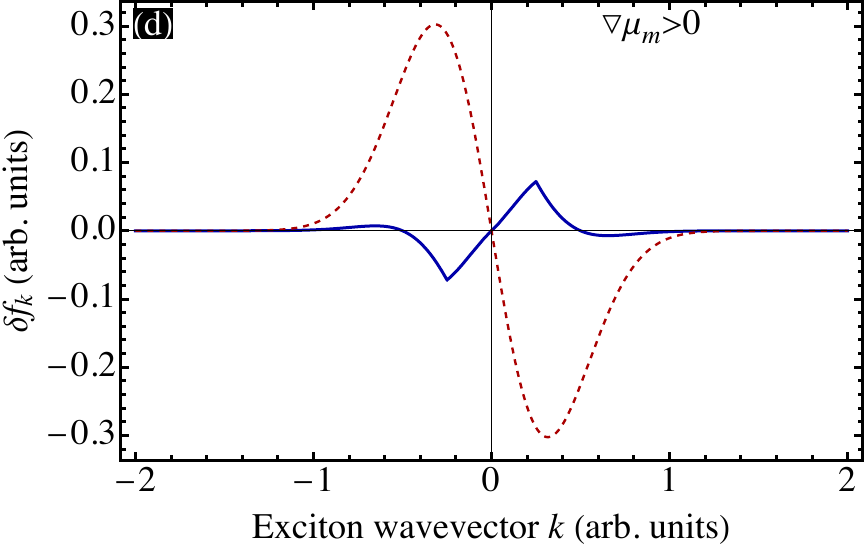}
	\caption{\textbf{1D toy model.} (a) Magnon dispersion $E_p = -\alpha|p| + p^2/2M_m$. (b) Magnon group velocity. 
		(c,d) Anisotropic part of exciton distribution $\delta f_k$ [blue solid curve, up to a positive prefactor] found numerically from Eq.~\eqref{delta:f:2:diff:1D} (solid curve) and using the small-$k$ asymptotics [red dotted curve, cf. Eq.~\eqref{delta:f:3:diff}]. Exciton dispersion $\varepsilon_k = \hbar^2 k^2/2M_x$. 
		We use the dimensionless parameters $\alpha=1$, $M_m=1$, $M_x/M_m = 0.1$, $T=0.3$. No cut-off ($p_0\to \infty$) in (c). Cut-off parameter $p_0=0.5$ in (d).}\label{1D:toy}
\end{figure}

Figure~\ref{1D:toy} shows the toy model dispersion of magnons (a), their group velocity (b) and the exciton distribution function found numerically from~\eqref{delta:f:2:diff:1D} (c,d) for two models of interaction. For panel (c) we consider momentum independent matrix element $M$. For panel (d) we consider a more general case with the cut-off momentum $p_0$:
\begin{equation}
\label{M:cut}
M(k,p,q) = \begin{cases}
M, \quad p<p_0,\\
0, \quad p \geqslant p_0.
\end{cases}
\end{equation}
In this model the interaction with magnons with large momenta is absent.
The result of numerical calculation of $\delta f(k)$ after Eq.~\eqref{delta:f:2:diff:1D} is shown in Fig.~\ref{1D:toy}(d). In contrast to Fig.~\ref{1D:toy}(c) [and large $p$/small $k$ asymptotics~\eqref{delta:f:3:diff} shown by the dashed curve] here the occupancies of excitons with $k>0$ are higher than those with $k<0$ and excitons flow counter to the magnon flux $\bm v \uparrow\downarrow \bm u$. The situation shown in panel (d) corresponds to the negative drag of excitons by magnons with
\[
\bm v \uparrow\downarrow \bm u.
\]

\section{Emission dynamics at different energies in the 10L crystal.}

\begin{figure*}[h!]
	\includegraphics[scale=1.0]{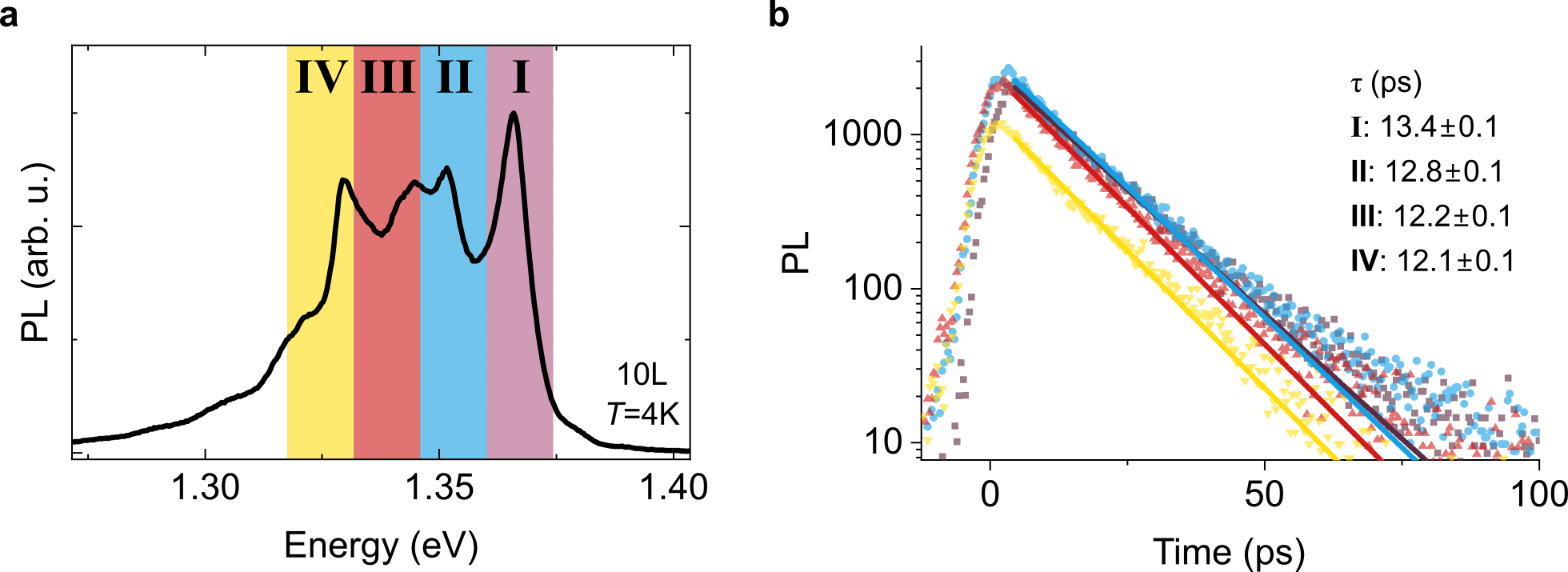}
	\caption{\textbf{Emission dynamics at different energies in the 10L PL signal.} 
		\textbf{a}~Time-integrated PL emission spectrum. 
		\textbf{b}~PL decay curves extracted from a spectrally resolved Streak camera measurement at the energy windows color-coded in a. Laser excitation fluence was 490 \SI{}{\micro\J\per\square\cm} and the sample temperature $T$ = 4\,K for all measurements. 
		\label{fig:SI-Fig-R3_1}}
\end{figure*}

We measured the spectrally dispersed PL decay of our 10L crystal at low temperature where spectral features are well resolved. 
The color-coded rectangles superimposed on the time-integrated PL spectrum \cref{fig:SI-Fig-R3_1}a indicate the respective spectral positions of the PL decay curves in \textbf{b} extracted from the Streak camera image. 
A single exponential fit shows that all four PL decay curves are essentially characterized by the same lifetime ($\tau= (12 \pm 1)\,ps$). 
It does not depend on the emission energy and is the same whether we analyze the $X_0$ peak or any of the other low-energy peaks. 
These results therefore complement our analysis shown in Extended Data Fig. 1: Whether we isolate the $X_0$ emission peak or integrate over all low-energy emissions and the $X_0$ peak, the spatio-temporal dynamics and the extracted diffusion coefficients are essentially the same. 

\section{Size of the zero-delay PL emission spot.}

\begin{figure*}[h!]
	\includegraphics[scale=1.0]{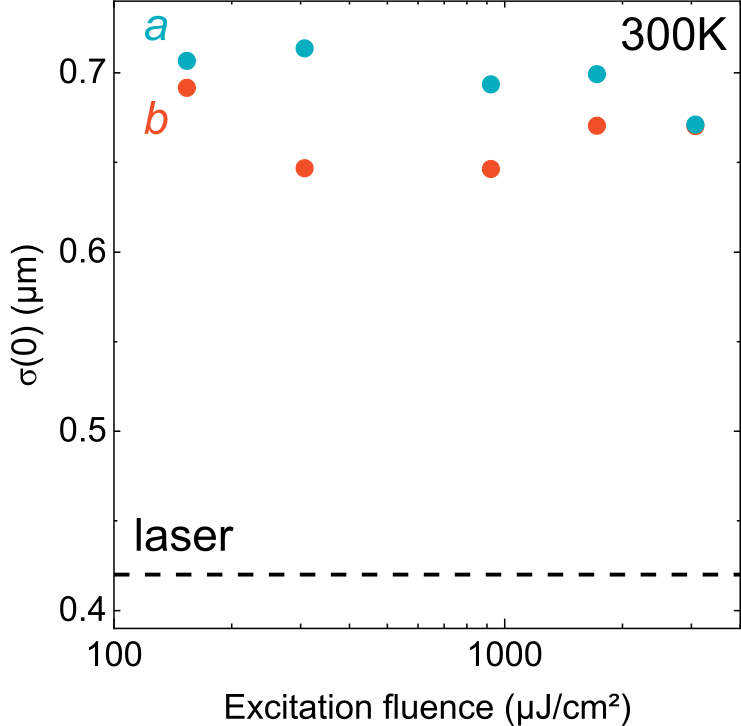}
	\caption{\textbf{Variance of the PL emission profile collected for four picoseconds.} 
	Data is shown for different crystallographic axes and excitation fluences. 
	The width of the laser imaged on the Streak camera is indicated by the black dashed line.
	\label{fig:SI-Fig-R2_1}}
\end{figure*}

\newpage
\section{Temperature dependence of exciton lifetimes}

Here, we briefly discuss the anomalous temperature dependence of the exciton lifetime shown in Extended Data Fig. 3b. 
In most materials, the emission lifetime of excitons tends to decrease with increasing temperatures, mainly due to faster, temperature-activated non-radiative recombination channels. 
Interestingly, in an ideal semiconductor quantum well with intrinsic radiative recombination, the lifetime is in fact expected to increase with temperature~\cite{Andreani1991}, as the excitonic distribution spreads out in momentum space, which reduces the fraction of excitons with momenta inside the light cone. 
Similar behavior is predicted for 2D semiconductors, with values in the range of 10’s to 100’s of ps for monolayer transition-metal dichalcogenides in the similar temperature range~\cite{Wang2016}.

For the magnetic semiconductor CrSBr, additional effects may be considered. 
A recent spectroscopy study reports an anomalous temperature dependent reduction of the exciton oscillator strength~\cite{Shao2025}.
The oscillator strength is observed to sizably decrease as a function of temperature, an unusual behavior when compared to Wannier-Mott excitons in other inorganic semiconductors. 
Calculations based on ab-initio many-body perturbation theory suggest an intriguing microscopic reason for this peculiar change in oscillator strength, which is related to the presence of spin fluctuations. 
Spin disorder modifies the orbital composition of the exciton wave function. 
More specifically, it quenches the inter-site contribution, i.e., the hopping of electrons and holes between neighboring Cr-atoms, which is to a large extent responsible for the large oscillator strength of excitons in CrSBr. 
Phenomenologically, it can be argued that magnetic disorder, by reducing inter-site hopping, leads to a decrease of the exciton oscillator strength, and, as a results, to an increase of the radiative lifetime of excitons. 
If the decay of the exciton population observed in our experiments is indeed limited by radiative recombination, this effect could qualitatively explain the peculiar dependence of the exciton lifetime in Extended Data Fig. 3b.

\clearpage
%